	\documentclass[a4paper,fleqn,usenatbib,useAMS]{mnras}

	\usepackage{graphicx}   
	\usepackage{amsmath}    
	\usepackage{amssymb}    
	\usepackage{multicol}        
	\usepackage{bm}         
	\usepackage{pdflscape}  
	\usepackage{footmisc}

	\usepackage[T1]{fontenc}
	\usepackage{ae,aecompl}

	\title[Czernik 3]{{  The} disintegrating old open cluster Czernik 3}

	\author[Sharma et al.]{
	Saurabh Sharma,$^{1}$\thanks{E-mail: saurabh@aries.res.in}
	Arpan Ghosh,$^{1}$, D. K. Ojha,$^{2}$  R. Pandey,$^{1}$ T Sinha,$^{1}$ 
	\newauthor
	A. K. Pandey,$^{1}$,  S. K. Ghosh,$^{2}$ N. Panwar,$^{1}$ and S. B. Pandey$^{1}$
	\\\\
	$^{1}$Aryabhatta Research Institute of Observational Sciences (ARIES), Manora Peak, Nainital, 263 002, India, saurabh@aries.res.in\\
	$^{2}$Tata Institute of Fundamental Research (TIFR), Homi Bhabha Road, Colaba, Mumbai - 400 005, India
	}

	\date{Accepted XXX. Received YYY; in original form ZZZ}

	\pubyear{2020}

\begin{document}
	\label{firstpage}
	\pagerange{\pageref{firstpage}--\pageref{lastpage}}
	\maketitle

	\begin{abstract}
	We have performed a detailed analysis of the Czernik 3 (Cz3) open cluster by using
	deep near-infrared photometry taken with TIRCAM2 on 3.6m Devasthal optical telescope
	along with the recently available high quality proper motion  data from the {\it Gaia} DR2
	and  deep photometric data from Pan-STARRS1.
	The cluster has a highly elongated morphology with fractal distribution of stars.
	The core and cluster radii of the cluster  are estimated as 0.5 pc and 1.2 pc, respectively.
	We have identified 45 stars as  cluster members using the {\it Gaia}  proper motion data.
	The distance and age of the cluster are found to be $3.5\pm0.9$ kpc and $0.9^{+0.3}_{-0.1}$ Gyr, respectively.
	The slope of the mass function  $`\Gamma'$  in the cluster region, in the mass range 
	$\sim$0.95$<$M/M$_\odot$$<$2.2, is found to be $-1.01\pm0.43$. 
	The cluster shows the signatures of mass-segregation and is dynamically relaxed (dynamical age=10 Myr). 
	This along with its small size,  big tidal radius, low density/large separation of stars, 
	and elongated and distorted morphology, indicate that the Cz3 is
	a loosely bound disintegrating cluster under the influence of external tidal interactions.
	\end{abstract}



	\section{Introduction}
	\label{sect:intro}

	Most of the stars form in a clustered
	environment in molecular clouds \citep{2003ARA&A..41...57L}.
	The dynamics of stars in the clusters as
	well as the structure of clusters measured as a function of cluster age 
	hold important clues on the processes of star formation and stellar
	evolution.  The Galactic disk is abundant in stars, spiral arms, and giant
	molecular clouds. Therefore, star clusters located in the disk are
	subjected to disturbance, such as disk shock, spiral arm passage,
	molecular cloud encounters, etc. \citep{1958ApJ...127..544S,2012MNRAS.426.3008K}.
	As clusters age, the expulsion of gas by stellar feedback
	as well as dynamical interactions between stars and binary systems
	in the cluster soften its gravitational potential, leading to their expansion and 
	to their partial or total dissolution into the field of their host galaxy 
	\citep[e.g.,][]{1958ApJ...127..544S,2011MNRAS.415.3439D,2012MNRAS.427..637P,2013MNRAS.436.3727D,2013MNRAS.432..986P,2013A&A...559A..38P,2017A&A...600A..49B,2018MNRAS.473..849D}.
	The typical survival timescale of open clusters in the Galactic
	disk is about 200 Myr \citep{2006A&A...446..121B,2013ApJ...762....3Y}.
	Open clusters much older than the survival timescale usually have
	distorted shape and loosened structure which leads to their
	disruption. The disintegrated open clusters will then become
	moving groups and supply field stars \citep{2019ApJ...877...12T}. 
	
	The core (nucleus) and the corona (extended region of the star cluster)
	are two main regions in open clusters \citep[e.g.,][]{1969SvA....12..625K,1990AJ.....99..617P,2006AJ....132.1669S}.
	The nucleus of the cluster usually contains bright and massive
	stars along with few low-mass stars, whereas
	the corona, relatively contains a large number of faint low-mass stars 
	\citep[][]{1999A&A...352L..69B,2006AJ....132.1669S,2008AJ....135.1934S}.
	Whether this segregation of massive stars towards the central region of a cluster
	occurs due to an evolutionary effect or is of primordial origin is not
	yet entirely clear. In the first scenario, massive stars may form anywhere
	in the cluster and eventually sink to the cluster center
	through the effects 
	of two-body relaxation \citep[e.g.,][]{2007ApJ...655L..45M,2009ApJ...700L..99A}.
	This is supported by numerical
	simulations in which mass segregation occurs on time-scales that
	are of the order of the cluster's ages \citep[e.g.,][]{2010MNRAS.407.1098A,2014MNRAS.445.4037P}. 
	The second scenario suggests that massive stars form preferentially
	in the central region of the cluster either by efficiently accreting gas
	due to their location at the bottom of the cluster potential well \citep[][]{2010ApJ...723..425D}
	or by a coalescence process of less massive stars
	\citep{2007JKAS...40..157D,2008ApJ...678L.105D}.
	The time-scale for this mass-segregation to complete is still not very well known and
	is important to know its implication on the dissolution of star clusters.
	The second scenario is an active area of research,
	especially because of the need to understand trapezium-type sub-systems in star clusters \citep{2000ASPC..198..105M},
	and the associated implications for the formation mechanisms of massive stars \citep{1998MNRAS.298...93B}.

\begin{figure*}
	\centering
	\hbox{
	\hspace{-0.5cm}
	\includegraphics[width=0.7\textwidth, angle=0]{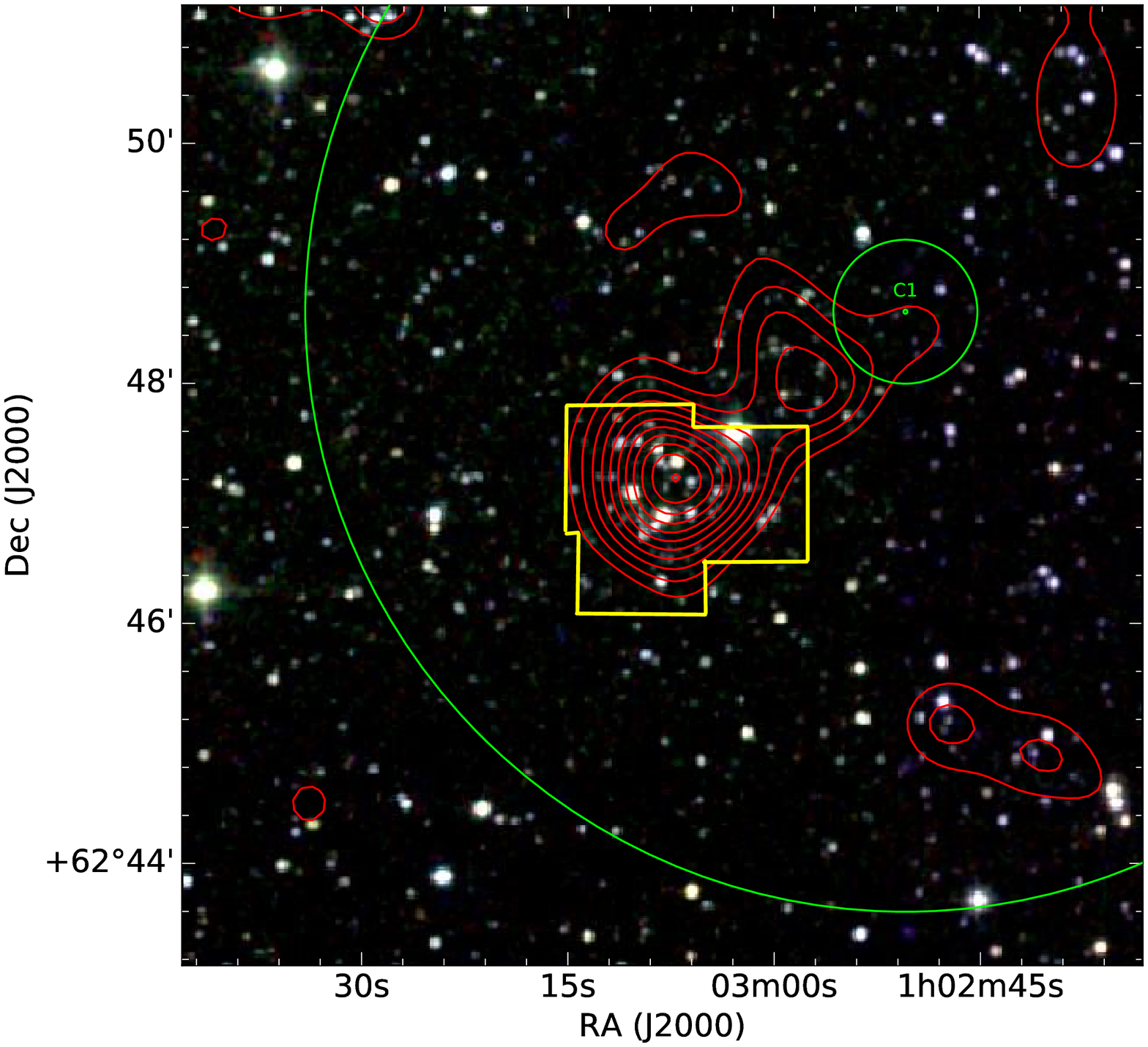}
	\hspace{-6.5cm}
	{\vbox 
	{
	\includegraphics[width=0.3\textwidth, angle=0]{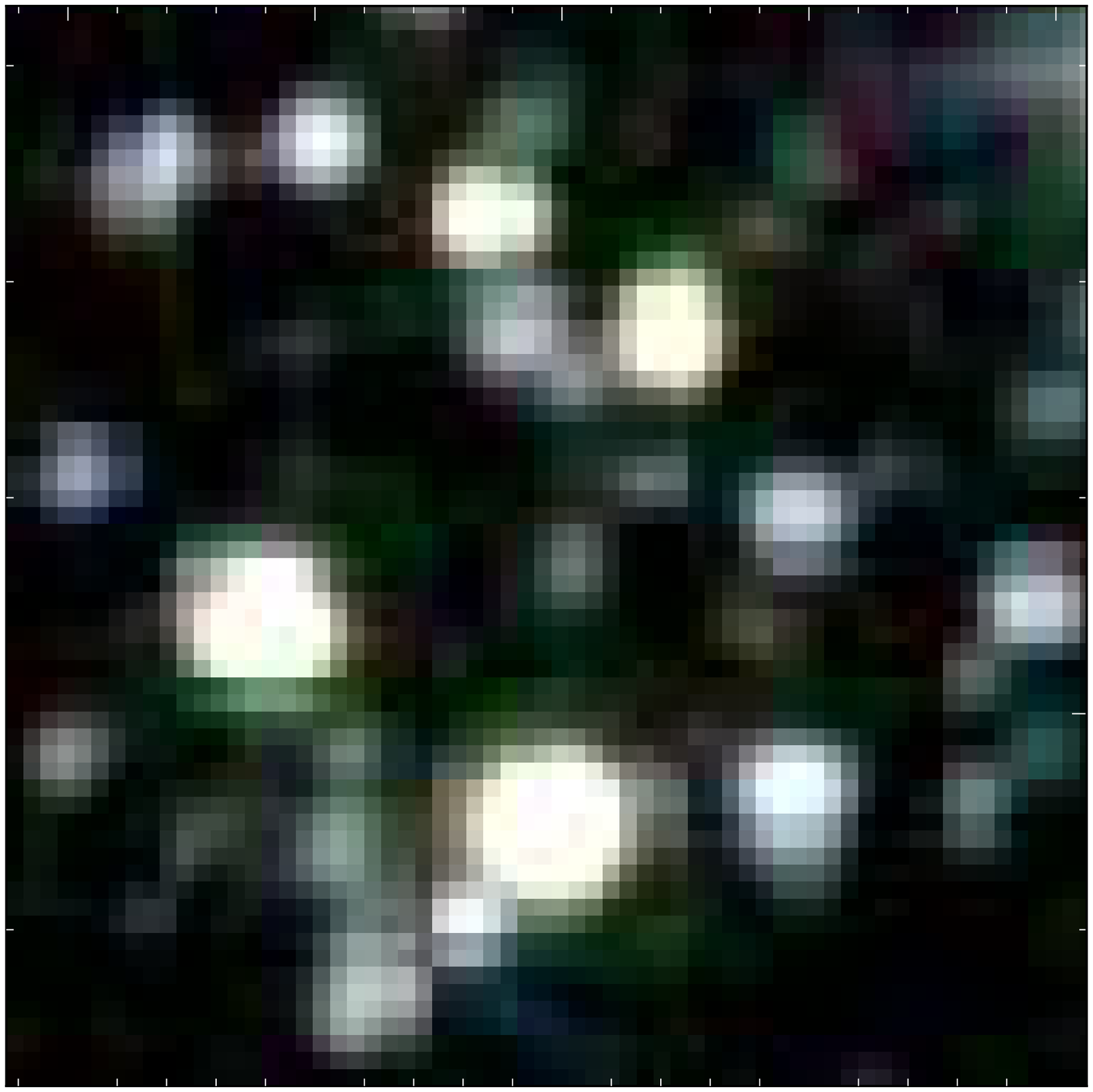}
	{\vbox{
	\vspace{0.6cm}
	\includegraphics[width=0.3\textwidth, angle=0]{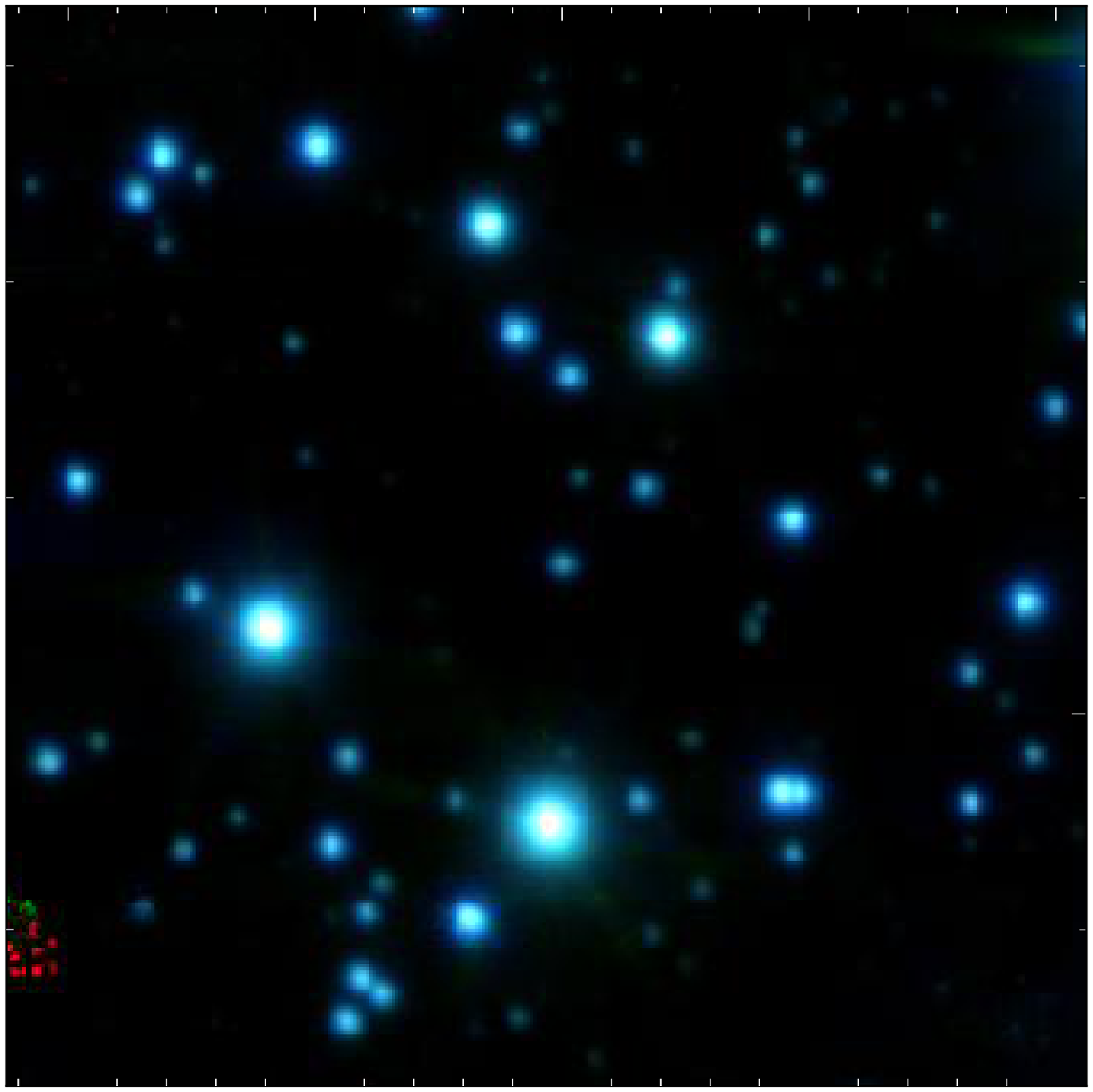}
	}}}}}
	\caption {
	(Left panel): Color-composite image obtained using the $J$ (blue), $H$ (green),
	and $K$ (red) 2MASS images for an area of $\sim10\times 10$ arcmin$^2$ around the Cz3 cluster.
	The yellow region encloses the present observations using TIRCAM2. The small 
	and big green circles are the core and the cluster regions identified by \citet{2017NewA...52...55B}. 
	The red contours are the isodensity contours generated using nearest 
	neighbor method from the 2MASS data (cf. Section 3.1).
	(Right panels): Comparison of the colour-composite images obtained by using the $J$ (blue), $H$ (green),
	and $K$ (red) images for the core region of the Cz3 cluster from the 2MASS (top panel) 
	and TIRCAM2 observations (bottom panel).
	}
	\label{2c}
	\end{figure*}

	During the past decade several studies on star clusters, 
	stellar evolution and dynamics have been carried out.
	However, most of  these studies are not always based on deep photometric data 
	and  lacks the membership determination based on high-quality proper motion (PM) data.
	Czernik 3 ($\alpha_{J2000}$: $01^h03^m06^s$, $\delta_{J2000}$: $+62^\circ47^\prime00^{\prime \prime}$ \citep{2002A&A...389..871D}; hereafter Cz3, cf.  Fig. \ref{2c}),
	one of the poorly studied open clusters, is located in the Galactic plane towards 
	the 2$^{nd}$ Galactic quadrant ($l$ = 124$^\circ$.265, $b$ = -0$^\circ$.058).
	This cluster is cataloged in the OPENCLUST\footnote{New Optically Visible Open Clusters and Candidates Catalog \citep{2002A&A...389..871D} \label{OPENCLUST}}
	as well as in MWSC\footnote{Milky Way Stellar Clusters catalog 
	\citep{2012A&A...543A.156K,2013A&A...558A..53K,2014A&A...568A..51S,2015A&A...581A..39S}\label{MWSC}} with  the distance and age in the 
	range 1.4 - 1.6 kpc and 100 - 630 Myr, respectively.
	Recently, \citet{2017NewA...52...55B} placed this  cluster at a distance of 1.75 kpc with an age of 115 Myr, using the 2MASS data.
	Since most of the information about this cluster is derived from the not-so-deep photometric surveys, we have
	revisited this cluster and performed a detailed analysis to understand its dynamical evolution
	by using our deep near-infrared (NIR) observations taken from the
	recently installed 3.6m telescope at Devasthal, Nainital, India \citep{2018BSRSL..87...29K}, along with
	the recently available data from  the {\it Gaia} data release 2 \citep{2016A&A...595A...1G,2018A&A...616A...1G}
	and Pan-STARR1 \citep{2016arXiv161205560C}.

	In this paper, Section \ref{sect:obs} describes the observations and data reduction.
	The structure of this cluster, membership probability of stars in the cluster region, 
	fundamental parameters (i.e., age and distance) of the cluster, and mass function (MF) analyses
	are presented in Section \ref{sect:result}.
	The dynamical structure of this cluster is discussed in Section \ref{sect:diss}, and
	we conclude our studies in Section \ref{sect:conclusion}.

	\section{Multi-wavelength  data sets}
	\label{sect:obs}

	\subsection{Deep NIR data}

	Deep NIR  $J$ (1.20 $\mu$m), $H$ (1.65 $\mu$m) and
	$K$ (2.19 $\mu$m) band photometric observations of the Cz3 open cluster (cf. Fig. \ref{2c}) 
	were taken during the night of 2017 October 7 using
	TIFR Near Infrared Imaging Camera-II \citep[TIRCAM2;][]{2012BASI...40..531N,2018JAI.....750003B}  mounted at the 
	 Cassegrain main port of the 3.6m ARIES Devasthal Optical Telescope \citep[DOT;][]{2018BSRSL..87...29K}. 
	The weather conditions in these nights were good with relative humidity $<60$ percent
	and the full-width at half maxima of the stellar images was $\sim$0.6 arcsec.
	The field of view (FOV) of the TIRCAM2 is $\sim86.5\times86.5$ arcsec square with a
	plate scale of 0.169 arcsec.
	We took four pointings of the Cz3 open cluster covering the central cluster region ($\sim$3 arcmin square)
	and in each pointing  several frames in a 5-point dither pattern were taken 
	with exposure times of 50 sec, 50 sec, and 10 sec in $J,~H,$ and $K$ bands, respectively. 
	The total exposure times were 500 sec, 500 sec and 1000 sec in each pointing in $J,~H,$ and $K$ bands, respectively. 
	Dark and sky flats were also taken during the observations.
	Sky frames in each filter were generated from the median of the dithered frames.

\begin{figure*}
	\centering
	\includegraphics[width=0.9\textwidth, angle=0]{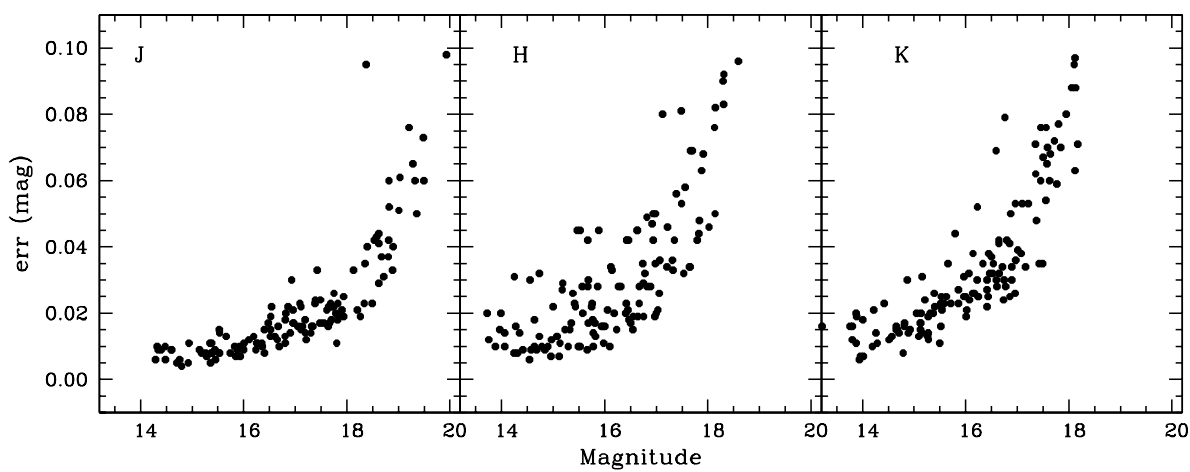}
	\caption {DAOPHOT errors as a function of $J,H,$ and $K$ magnitudes.}
	\label{fig:calibrate}
	\end{figure*}

	The basic image processing such as dark/sky subtraction and flat fielding was done using tasks 
	available within IRAF\footnote{IRAF is distributed by the National Optical Astronomy Observatory, 
	which is operated by the Association of Universities for Research in Astronomy (AURA) 
	under cooperative agreement with the National Science Foundation.}.
	Instrumental magnitudes were obtained  using the DAOPHOT package. 
	As the cluster region is  crowded we carried out point spread function (PSF) photometry to get the magnitudes of the stars. 
	Astrometry of the stars was done using the Graphical Astronomy and Image 
	Analysis Tool\footnote{http://star-www.dur.ac.uk/~pdraper/gaia/gaia.html} with a rms noise of the order of $\sim$0.1 arcsec.
	The calibration of the photometry to the
	standard system was done using the following transformation equations:

	\begin{equation}
	(J-H)= (0.84\pm 0.13)\times(j-h)+ (0.35\pm 0.03)
	\end{equation}
	\begin{equation}
	(J-K)= (0.77\pm 0.12)\times(j-k)+ (1.11\pm 0.07)
	\end{equation}
	\begin{equation}
	(J-j)= (-0.12\pm 0.13)\times(J-H)- (4.06\pm 0.07)
	\end{equation}

	where the capital $JHK$ are the standard magnitudes of the
	stars taken from the 2MASS catalog and  the 
	small $jhk$ are the present instrumental magnitudes of the similar stars normalized per sec exposure time.
	The typical DAOPHOT errors as a function of corresponding standard magnitudes are
	shown in Fig. \ref{fig:calibrate}. 
	We have used only those stars for further analyses which are having
	signal-to-noise ratio greater than 10 (photometric errors $<$ 0.1 mag).
	In total, 133 stars were identified in the central region of the Cz3 cluster with detection limits of
	19.9 mag, 18.6 mag, 18.2 mag in $J$, $H$, $K$ bands, respectively.
	Some of the brighter stars (18 in total) were saturated in our observations; we have taken their respective magnitudes from
	the 2MASS point source catalog. The positions and magnitudes of the stars in different bands
	are given in Table \ref{table1}.

	\begin{table*}
	\centering
	\caption{\label{Tyso} Sample of the NIR photometry of the stars in the present study. The complete table is available in the electronic form only.}
	\begin{tabular}{@{}l@{ }c@{ }c@{ }r@{ }r@{ }r@{ }}
	\hline
	ID& $\alpha_{(2000)}$&$\delta_{(2000)}$& $J\pm\sigma~~~~~$&$H\pm\sigma~~~~~$&$K\pm\sigma~~~~~$\\
	& {\rm $(degree)$} & {\rm $(degree) $} & (mag)$~~~~~$&  (mag)$~~~~~$& (mag)$~~~~~$ \\
	\hline
	1  &    15.762209 &     +62.793190  &  $ 8.523 \pm 0.021$&$ 7.937 \pm 0.017$&$ 7.787\pm  0.018$ \\
	2  &    15.783788 &     +62.781624  &  $10.577 \pm 0.024$&$ 9.600 \pm 0.021$&$ 9.277\pm  0.020$ \\
	3  &    15.793315 &     +62.784641  &  $10.904 \pm 0.021$&$10.092 \pm 0.017$&$ 9.866\pm  0.019$ \\
	4  &    15.779851 &     +62.789120  &  $12.475 \pm 0.025$&$11.515 \pm 0.024$&$11.127\pm  0.022$ \\
	5  &    15.784626 &     +62.772507  &  $12.550 \pm 0.021$&$12.047 \pm 0.021$&$11.940\pm  0.023$ \\
	\hline
	\end{tabular}
	\label{table1}
	\end{table*}

	\subsection{Archival data}
	\label{sect:obs1}

	In order to study a wider area around Cz3, we have selected a FOV of $10\times10$ arcmin square
	as shown in Fig. \ref{2c} and downloaded the available data from the below mentioned archives:

	\subsubsection{Gaia}
	The second {\it Gaia} data release, {\it Gaia} DR2 \citep{2016A&A...595A...1G,2018A&A...616A...1G}, which consists 
	of astrometry, photometry, radial velocities, and information on astrophysical parameters and 
	variability for sources brighter than  $G$ (0.33-1.05 $\micron$) = 21 magnitude.
	{\it Gaia} DR2 contains celestial positions and apparent brightness in $G$ band for approximately 1.7 billion sources.
	In addition, for 1.3 billion of those sources, parallaxes and PMs are available.
	This data release also contains the broad-band colour information in the form of the 
	apparent brightness in the $G_{BP}$ (0.33-0.68 $\micron$) and $G_{RP}$ (0.63-1.05 $\micron$) bands for 1.4 billion sources.
	We have downloaded the {\it Gaia}  DR2 data from the data archive\footnote{https://gea.esac.esa.int/archive/}.
	The $Gaia$ sample consists of stars with error in their PM values less than 3 mas yr$^{-1}$.

	\subsubsection{Pan-STARRS1}
	The Panoramic Survey Telescope and Rapid
	Response System (Pan-STARRS1 or PS1) imaged the entire northern sky, above a declination
	of $\delta \sim -30^\circ$, in five broadband filters $(g,r,i,z,y)_{P1}$, which
	encompass the spectral window from $\sim$0.4 - 1.05 $\micron$. The
	survey, which included multiple pointings per observed field per
	filter, reached a depth of $g_{P1} \sim 23.2$ mag,
	$r_{P1} \sim 23.2$ mag, and $i_{P1} \sim 23.1$ mag in the stacked images.
	The details about the PS1 surveys and latest data products are given in \citet{2016arXiv161205560C}.
	We have downloaded images\footnote{http://ps1images.stsci.edu/cgi-bin/ps1cutouts} as well as
	point source catalog\footnote{http://catalogs.mast.stsci.edu/} from  the data release 2 of the PS1.
	While PS1 had observations at multiple epochs, 
	however the analysis reported in the ensuing sections have used  average magnitudes in each band.
	The PS1 sample consists of stars  with photometric uncertainties less than 0.1 mag in different PS1 bands.

	\subsubsection{2MASS}
	We have also used the 2MASS point source catalog \citep{2003yCat.2246....0C} for NIR (JHK$_s$)
	photometry in the Cz3 region. This catalog is reported to be 99 percent complete down
	to the limiting magnitudes of 15.8 mag, 15.1 mag and 14.3 mag in the $J$ (1.24 $\mu$m), $H$ (1.66 $\mu$m) and
	$K_s$ (2.16 $\mu$m) band, respectively\footnote{http://tdc-www.harvard.edu/catalogs/tmpsc.html}.
	The 2MASS sample consists of stars with photometric uncertainties less than 0.1 mag in the $J$, $H$ and $K_s$ bands.

	\section{Results and Analysis}
	\label{sect:result}

	\subsection{Structure of  the Cz3 cluster}

\begin{figure*}
	\hspace{-0.55 cm}
	\includegraphics[width=0.495\textwidth, angle=0]{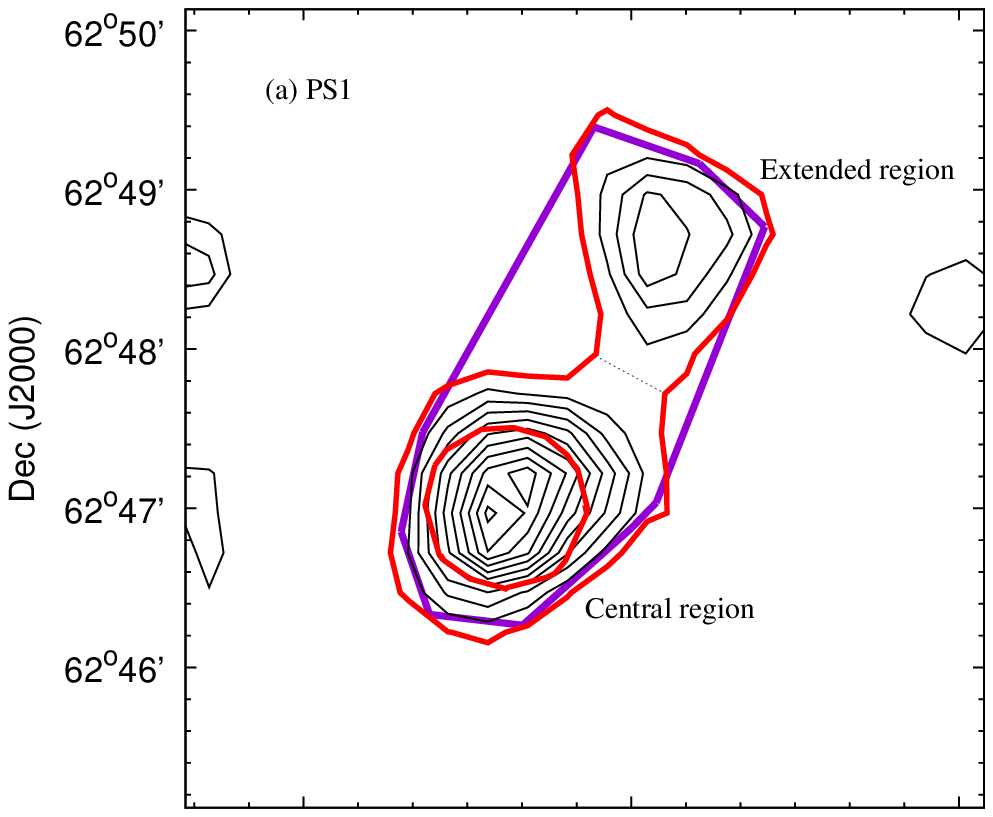}{
	\hspace{0.54cm}
	\includegraphics[width=0.405\textwidth, angle=0]{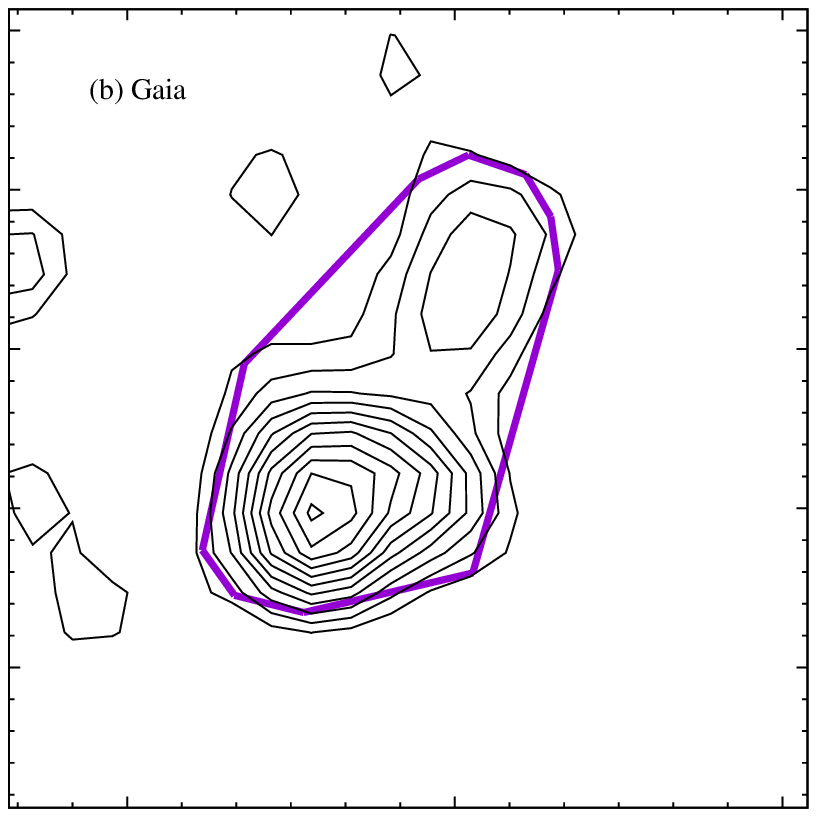}}
	\includegraphics[width=0.54\textwidth, angle=0]{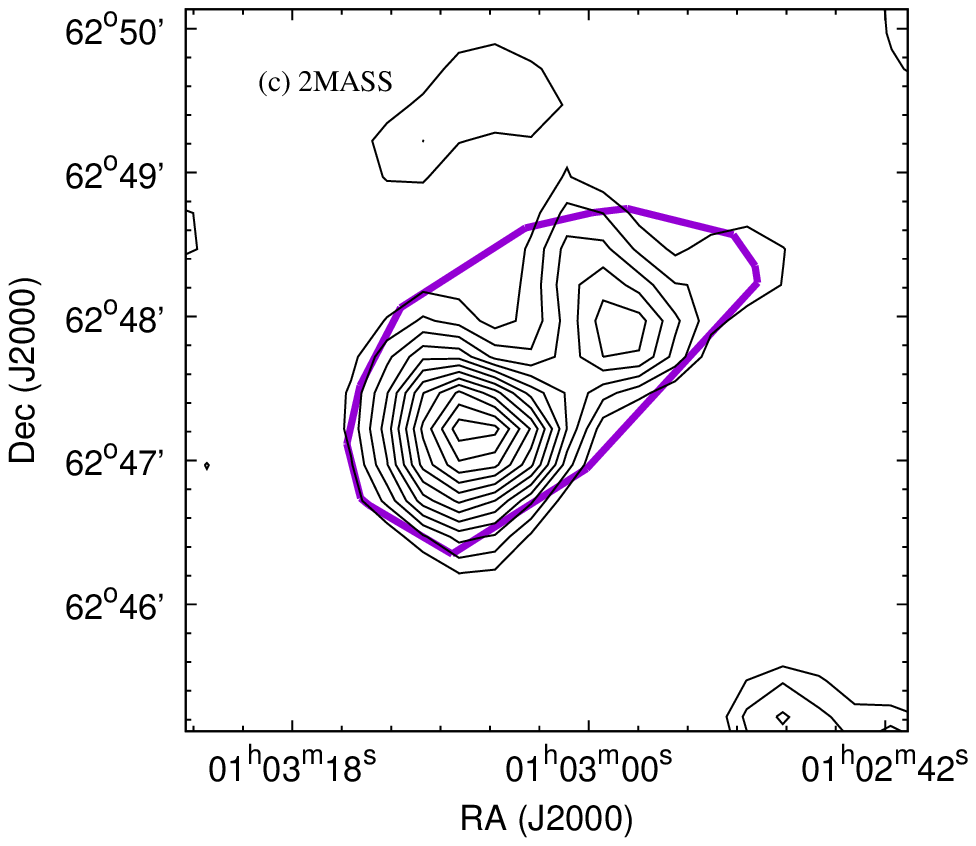}
	\includegraphics[width=0.44\textwidth, angle=0]{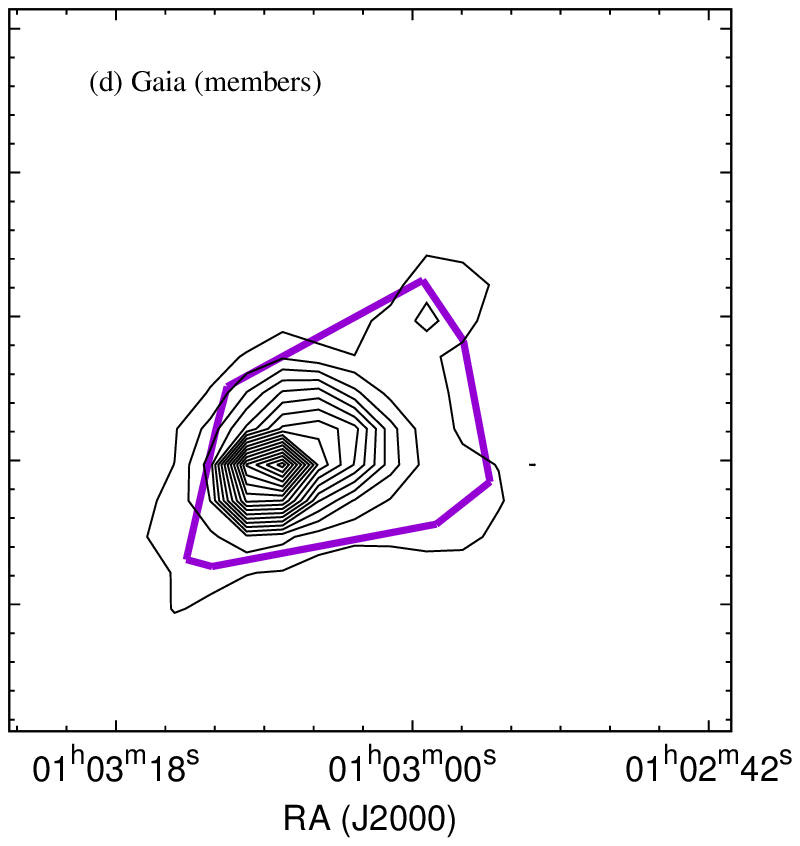}
	\caption {  Stellar density contours (black curves) generated by using the nearest neighbor method from the PS1 (panel a), $Gaia$ (panels b and d), and 2MASS (panel c) 
	data.  Panel (d) represents isodensity contours derived by using the location of member stars (cf. Section 3.2).
	The purple curve represents the Qhull for the stars located within the lowest density contour.
	In panel (a), the red density contour and red circle represent the cluster and core regions, respectively. 
	The dotted line represents the demarcation between the central and extended regions.
	}
	\label{fig:qhull}
	\end{figure*}

	To study the structure of the Cz3 open cluster, we obtained stellar number density maps
	for the sample of stars {  taken from the PS1, $Gaia$, and 2MASS surveys covering $10\times10$ arcmin 
	square FOV around this cluster region. We have also generated the stellar number density map for a sample of 
	member stars of Cz3 cluster (cf. Section 3.2).} The stellar number density maps were generated using the nearest neighbor (NN) method as
	described by  \citet{2005ApJ...632..397G}. We took the radial
	distance necessary to encompass the sixth nearest stars and
	computed the local surface density in a grid size of 5 arcsec \citep[cf.][]{2009ApJS..184...18G}. 
	The stellar number density contours derived by this method are plotted in Fig. \ref{fig:qhull} as black curves
	smoothened to a grid of size $3\times3$ pixels.
	The lowest contour is 1$\sigma$ above the mean of stellar density  
	and the step size is equal to the 1$\sigma$.
	As can be seen from the contours, the central core region of this cluster is almost circular,
	whereas the outer region is highly elongated and has another  peak of lower density.
	{  
	To study the structure of the cluster, a reliable membership, e.g., on the basis of PM,  is required. 
	The PM errors in the Gaia data towards fainter magnitudes are very high, hence we used the PS1 data which has similar 
	depth as Gaia data to demarcate the cluster extent and core region of the Cz3. The cluster  extent (lowest density contour) 
	and core (region within the half of the peak value of stellar density) are shown with red circle and red contour,
	respectively, in the panel (a) of Fig. \ref{fig:qhull}. 
	We have divided the cluster region into two parts, one containing the central region 
	and other containing the extended region. 
	The demarcation between these two regions is shown with a dotted line.} 
	The core region,  centered at $\alpha_{J2000}$: $01^h03^m06^s.9$, $\delta_{J2000}$: $+62^\circ47^\prime00^{\prime \prime}$, 
	has a  radius of $\sim$30 arcsec.
	\citet{2017NewA...52...55B} have derived slightly off-center coordinates of 
	this cluster as  $\alpha_{J2000}$: $01^h02^m50^s.4$, $\delta_{J2000}$: $+62^\circ48^\prime36^{\prime \prime}$,
	with core and cluster radii as 36 arcsec and 300 arcsec, respectively (cf. Fig. \ref{2c}).

        \begin{table*}
        \centering
        \caption{\label{struc} Structural parameters of the CZ3 cluster.}
        \begin{tabular}{@{ }c@{ }c@{ }r@{ }c@{ }c@{ }c@{ }c@{ }c@{ }c@{ }}
        \hline
        ID&  Sample & n$_{total}$ & $n_{hull}$ & A$_{hull}$ & A$_{cluster}$ & R$_{cluster}$ & R$_{circ}$ & Aspect  \\
	  &         &   &   &$(arcmin^2)$& $(arcmin^2)$  & $(arcmin)$    & $(arcmin)$ & ratio   \\
        \hline
	(a) & PS1     & 131& 9  & 4.05      & 4.35          &  1.18         & 1.65       &  1.96     \\
	(b) & $Gaia$  & 176& 10 & 4.16      & 4.40          &  1.18         & 1.61       &  1.86     \\
	(c) & 2MASS   & 73 & 13 & 4.00      & 4.87          &  1.25         & 1.60       &  1.64     \\
	(d) & Gaia (members) & 45 & 8  & 2.69      & 3.27          &  1.02         & 1.27       &  1.55     \\
        \hline
	  & Average & -  & -  &   3.73    & 4.22          &  1.16         & 1.53       &  1.75     \\
        \hline 
        \end{tabular}
        \end{table*}

	As the cluster is showing highly elongated morphology, we refine the definition of the cluster area using the convex 
	hull\footnote{Convex hull is a polygon enclosing all points in a grouping with internal angles between two
	contiguous sides of less than 180$^\circ$.} 
	rather than a circular area around the objects, as the later  method tends to significantly 
	overestimate the area of the cluster, in particular if the cluster is elongated or 
	irregularly shaped rather than spherical \citep{2006A&A...449..151S}.
	The convex hull for the stars located within the lowest density contour is computed 
	using the program Qhull\footnote{Barber, C. B., D.P. Dobkin, and H.T. Huhdanpaa, "The Quickhull Algorithm for Convex Hulls," ACM Transactions on Mathematical Software, 22(4):469-483, Dec 1996, www.qhull.org [http://portal.acm.org; http://citeseerx.ist.psu.edu].}.
	We estimate the area of the cluster $A_{cluster}$  using the convex hull of the
	data points, normalized by an additional geometrical factor taking into account 
	the ratio of the number of objects inside and on the convex hull \citep{1983ApJ...268..527H,2006A&A...449..151S}:

	\begin{equation}
	A_{cluster} = {{A_{hull}}\over{(1 - {n_{hull}\over n_{total}})}}
	\end{equation}

	where, $A_{hull}$ is the  area of
	convex hull, $n_{hull}$ is the number of vertices on convex hull, and $ n_{total}$ is the
	number of objects. The correction factor is used since the
	convex hull by itself tends to be smaller than the true sampling
	window. To be consistent, we define the cluster radius $R_{cluster}$ as the
	radius of a circle with the same area $A_{cluster}$.
	We have also estimated the circular radial size, $R_{circ}$, defined as half
	of the largest distance between any two members, the radius
	of the minimum area circle that encloses the entire grouping.
	{  For each data sets, the estimated  values for  $A_{hull}$,  $A_{cluster}$, $R_{cluster}$ and  $R_{circ}$
	are listed in Table \ref{struc}.
	The ratio $R^2_{circ}\over{R^2_{cluster}}$, which is a measure of 
	the aspect ratio of a cluster \citep{2009ApJS..184...18G} is also listed in the same table.
	In general, these values  suggest towards a highly elongated morphology of this cluster.
	The estimated extent of the cluster ($R_{cluster}$ =1.18 arcmin from PS1 sample) is much smaller than the values reported in the previous studies.}

	\subsection{Membership Probability}

	{\it Gaia} DR2  release has opened up the possibility of an entirely new perspective on the problem of 
	membership determination in cluster studies by
	providing the new and precise parallax measurements upto very faint limits\footnote{https://www.cosmos.esa.int/web/gaia/dr2}.
	To determine the membership probability, we adopted the method described in 
	\citet{1998A&AS..133..387B}. This method has previously been used extensively (see e.g., 
	\citet[][$\omega$ Centauri]{2009A&A...493..959B}; \citet[][NGC 6809]{2012A&A...543A..87S}; \citet[][NGC 3766]{2013MNRAS.430.3350Y}; 
	\citet[][NGC 3201]{2017AJ....153..134S}). 
	In this method, we first construct the frequency distributions of cluster
	stars ($\phi^\nu_c$) and field stars ($\phi^\nu_f$) using the following equations:

	\begin{equation}
	\begin{centering}
	\begin{tiny}
	\begin{split}
	\phi^\nu_c = {1\over{2\pi \sqrt{(\sigma^2_c + \epsilon^2_{xi})(\sigma^2_c + \epsilon^2_{yi})  }  } } \\ \times {\exp \left\{ {-{1\over2} \left [ \, { {{(\mu_{xi}-\mu_{xc})^2}\over{\sigma^2_c + \epsilon^2_{xi}}} + {{(\mu_{yi}-\mu_{yc})^2}\over{\sigma^2_c + \epsilon^2_{yi}}}  } \right ] \, } \right\}    }
	\end{split}
	\end{tiny}
	\end{centering}
	\end{equation}

	\begin{equation}
	\begin{centering}
	\begin{tiny}
	\begin{split}
	\phi^\nu_f = {1\over{2\pi \sqrt{1-\gamma^2} \sqrt{(\sigma^2_{xf} + \epsilon^2_{xi})(\sigma^2_{yf} + \epsilon^2_{yi})  }  } } \\ \times {\exp \left\{ {-{1\over{2(1-\gamma^2)}} \left [ \, { {{(\mu_{xi}-\mu_{xf})^2}\over{\sigma^2_{xf} + \epsilon^2_{xi}}}  - { {2\gamma(\mu_{xi}-\mu_{xf})(\mu_{yi}-\mu_{yf})}\over{\sqrt{(\sigma^2_{xf}+\epsilon^2_{xi})(\sigma^2_{yf}+\epsilon^2_{yi})}} } +  {{(\mu_{yi}-\mu_{yf})^2}\over{\sigma^2_{yf} + \epsilon^2_{yi}}} } \right ] \, } \right\}    }
	\end{split}
	\end{tiny}
	\end{centering}
	\end{equation}

	where $\mu_{xi}$, $\mu_{yi}$ and $\epsilon_{xi}, \epsilon_{yi}$ are 
	the PMs and  the errors in PMs of the $i^{th}$ star, respectively.
	$\mu_{xc}$, $\mu_{yc}$ and  $\sigma_c$, and $\mu_{xf}$, $\mu_{yf}$  and $\sigma_{xf}$, $\sigma_{yf}$
	 are the PM center and the PM dispersion of the cluster and field stars, respectively. 
	{  Subscript `x' refers to the RA component of the PM corrected with geometric transformation, 
	whereas subscript `y' refers to the Dec component of the PM.}
	$\gamma$ is the correlation coefficient which is calculated as:

	\begin{equation}
	\gamma =  { {(\mu_{xi} - \mu_{xf})(\mu_{yi} - \mu_{yf})}\over{\sigma_{xf} \sigma_{yf}}  }
	\end{equation}

	\begin{figure*}
	\centering
	\includegraphics[width=0.48\textwidth]{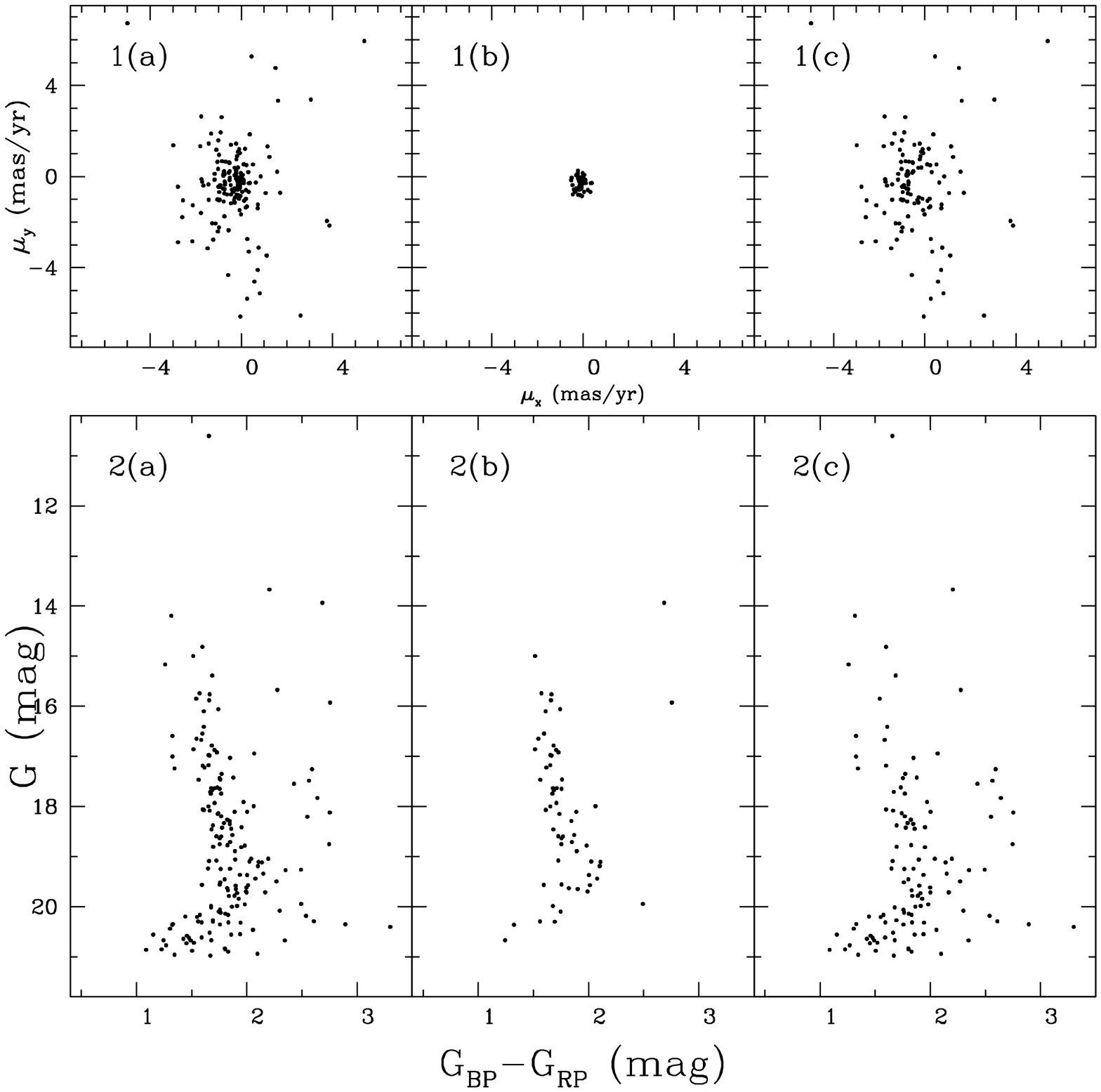}
	\includegraphics[width=0.48\textwidth]{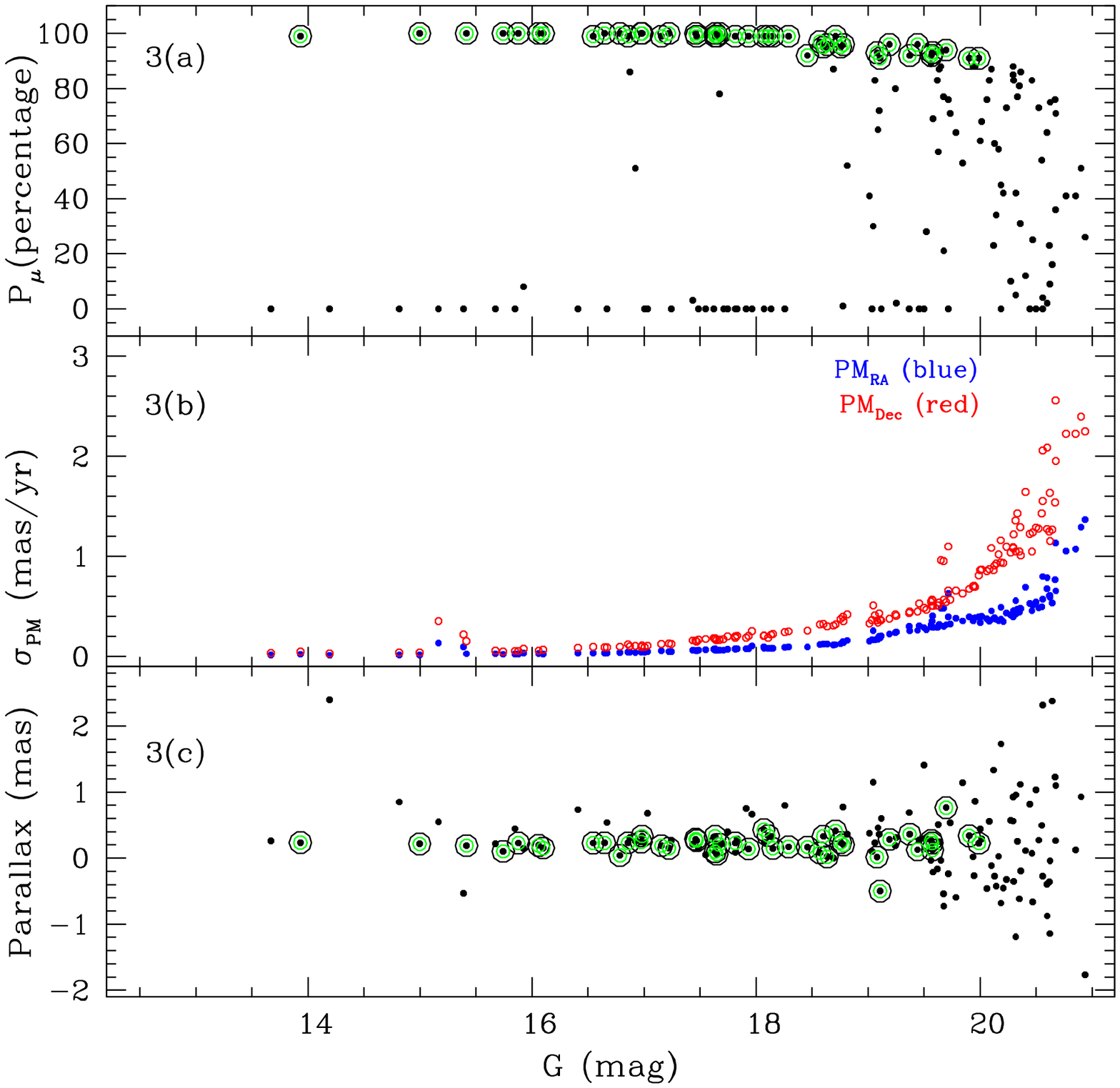}
		\caption{\label{pm1} Panel 1 \& 2: PM vector-point diagrams (panel 1) and
	{\it Gaia} DR2 $G$ vs. $(G_{BP} - G_{RP})$ CMDs (panel 2) for  stars located in the Cz3 cluster region.
	The left sub-panels (1(a) and 2(a)) show all stars, while the middle (1(b) and 2(b)) and the right sub-panels (1(c) and 2(c)) show
	the probable cluster members and field stars, respectively.
	Panel 3: Membership probability P$_\mu$, PM errors $\sigma_{PM}$ and parallax of stars as
	a function of $G$ magnitude for stars in the Cz3 cluster region.
	Forty five stars with P$_\mu>$90 percent are considered as member of the Cz3 cluster
	and  are shown by green circles with black rings.
	}
	\end{figure*}

	{\it Gaia}  PM data located within the convex hull of the cluster region
	are  used here to determine membership probability of cluster member stars.
	{  PMs $\mu_{x}$ and $\mu_{y}$  are} plotted as vector-point diagrams (VPDs) in the panel 1 of
	Fig. \ref{pm1}. The panel 2 shows the corresponding $G$ versus $G_{BP} - G_{RP}$
	colour-magnitude diagrams (CMDs). 
	{  The dots in sub-panel 1(a) represent PM distribution of all the stars 
	where a prominent  clump within a radius of $\sim$0.6 mas yr$^{-1}$ centered at  
	$\mu_{xc}$ = -0.23 mas yr$^{-1}$, $\mu_{yc}$ = -0.31 mas yr$^{-1}$ can bee seen. 
	This population of stars have almost similar PMs and have high probability for cluster membership.
	Remaining stars with scattered PM values are most probably a field population.
	This is more clearer in the VPDs and CMDs of probable cluster and field populations as shown in the
	sub-panels 1(b), 1(c), 2(b), and 2(c), respectively.
	The probable cluster members are showing well defined MS in the CMD which is usually seen for the similar population of stars.
	On the other hand, the probable field stars are quite obvious by their broad distribution in the CMD.}
	Assuming a distance of 3.5 kpc (the present estimate for the Cz3 cluster, Section 3.3) and a
	radial velocity dispersion of 1 km s$^{-1}$ for open clusters \citep{1989AJ.....98..227G},
	the expected dispersion ($\sigma_c$) in PMs would be $\sim$0.06 mas yr$^{-1}$.
	For field stars we have: $\mu_{xf}$ = -0.88 mas yr$^{-1}$, $\mu_{yf}$ = -0.58 mas yr$^{-1}$, $\sigma_{xf}$ = 5.9 mas yr$^{-1}$
	and $\sigma_{yf}$ = 3.2 mas yr$^{-1}$.
	The membership probability (ratio of distribution of cluster stars with all the stars) for the $i^{th}$ star is:

	\begin{equation}
	P_\mu(i) = {{n_c\times\phi^\nu_c(i)}\over{n_c\times\phi^\nu_c(i)+n_f\times\phi^\nu_f(i)}}
	\end{equation}

	where $n_c$ (=0.29) and $n_f$(=0.71) are the normalized number of stars for the cluster
	and field regions ($n_c$+$n_f$ = 1).

	The membership probability estimated as above, errors in the PM,
	and parallax values are plotted as a function of $G$ magnitude in
	panel 3 of Fig. \ref{pm1} . As can be seen in this plot, a high
	membership probability (P$_\mu >$ 90 percent) extends down to $G\sim$20 mag. 
	At brighter magnitudes, there is a clear separation between
	cluster members and field stars supporting the effectiveness of this
	technique. Errors in PMs become very high at faint limits, and the
	maximum probability gradually decreases at those levels. Except
	for a few outliers, most of the stars with high membership
	probability (P$_\mu >$ 90 percent) are following a tight distribution.
	Finally, on the basis of above analysis, 
	45 stars were assigned as cluster members based on their high membership probability P$_\mu$ ($>$90 percent;
	green circles with black rings in Fig. \ref{pm1} (panels 3a and 3c)).
	The details of these member stars are given in Table \ref{table2}.

	\begin{table*}
	\centering
	\caption{\label{Tyso} Sample of 45 stars identified as cluster members. The complete table is available in the electronic form only.}
	\begin{tabular}{ccrcccccc}
	\hline
	ID& $\alpha_{(2000)}$&$\delta_{(2000)}$& Parallax$\pm\sigma~~~~~$&$\mu_\alpha\pm\sigma$&$\mu_\delta\pm\sigma$ & $G$ & $G_{BP}-G_{RP}$ & Probability\\
	& {\rm $(degree)$} & {\rm $(degree) $} & (mas)&  (mas/yr)& (mas/yr) & (mag) & (mag) & (Percentage)\\
	\hline
	1&  15.800448& 62.774239&$   0.338\pm  0.454$&$  -0.424\pm  0.398 $&$ -0.389\pm  0.676 $&  19.903   & 2.135   &  91 \\
	2&  15.765736& 62.776619&$   0.764\pm  0.395$&$  -0.007\pm  0.318 $&$ -0.009\pm  0.545 $&  19.696   & 1.992   &  94 \\
	3&  15.782061& 62.777065&$   0.412\pm  0.192$&$  -0.182\pm  0.118 $&$ -0.155\pm  0.319 $&  18.709   & 1.849   &  99 \\

	\hline
	\end{tabular}
	\label{table2}
	\end{table*}

	\begin{figure}
	\centering
	\includegraphics[width=0.38\textwidth, angle=0]{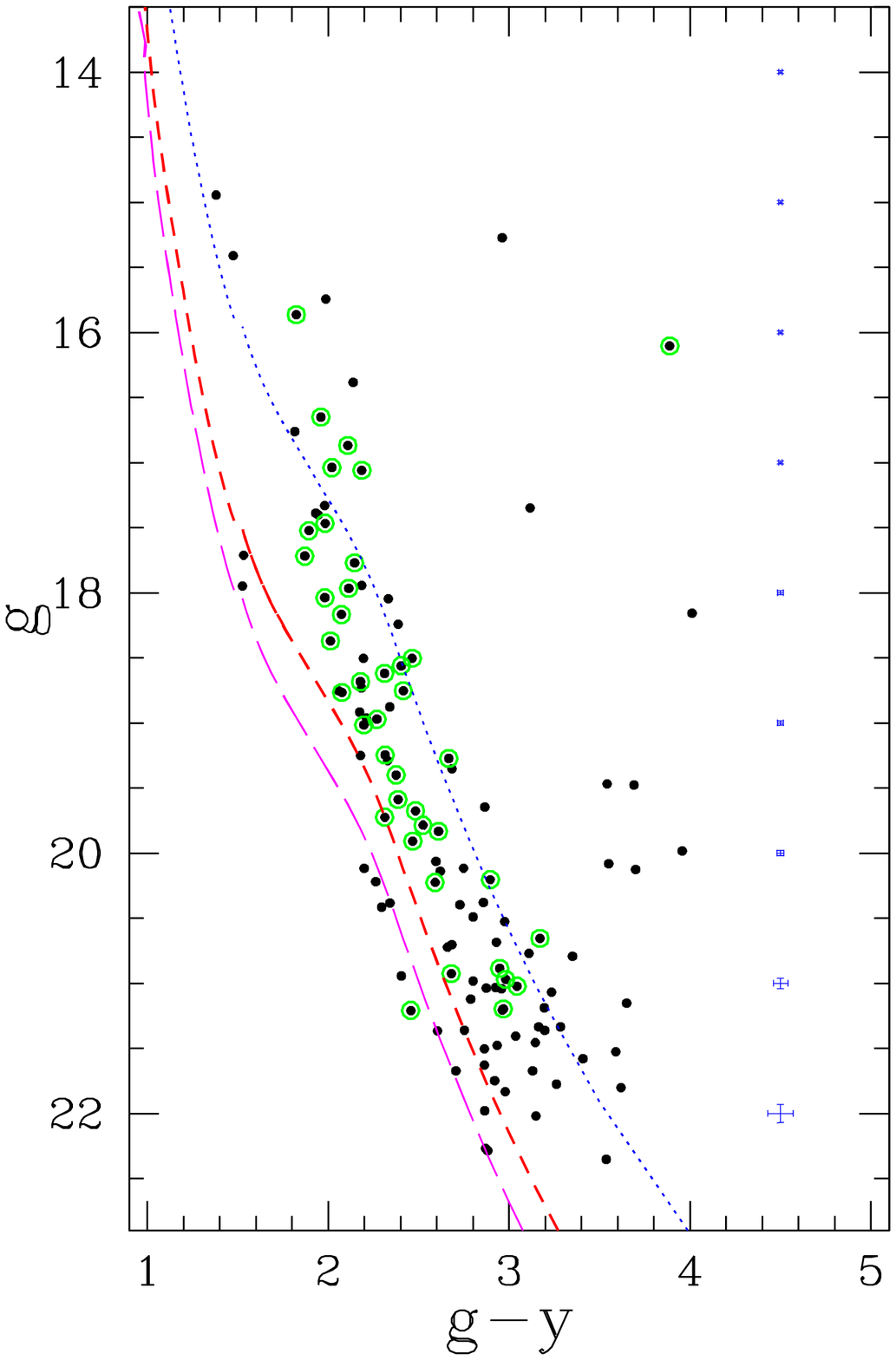}
	\caption {PS1 $g$ vs. $(g - y)$ CMD for the stars  in the cluster region (black dots).
	The green circles are the member stars of the cluster.
	The curves denote a ZAMS derived from \citet{2019MNRAS.485.5666P}
		corrected for extinction ($A_V$=2.9 mag) and distances of 1.75 kpc 
		\citep[blue-dotted curve,][]{2017NewA...52...55B},
		3.5 kpc (red-dashed curve, present analysis) and 4.5 kpc 
		\citep[purple-dashed curve,][]{2018yCat..36180093C}.
	{  Photometric error bars are also shown in the CMD.}
	}
	\label{fgy1}
	\end{figure}

	\begin{figure*}
	  \centering
	 \includegraphics[width=0.33\textwidth, angle=0]{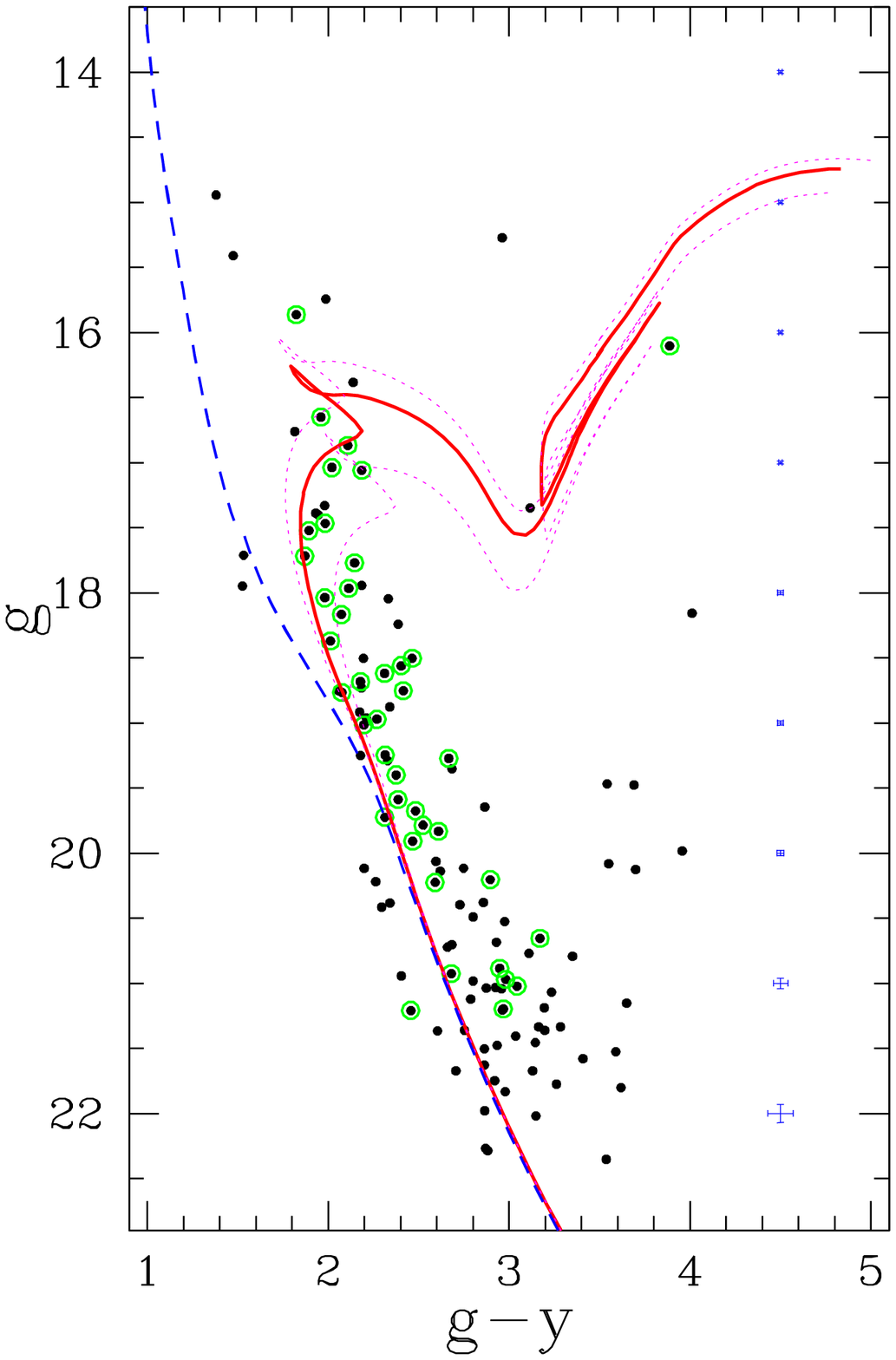}
	 \includegraphics[width=0.33\textwidth, angle=0]{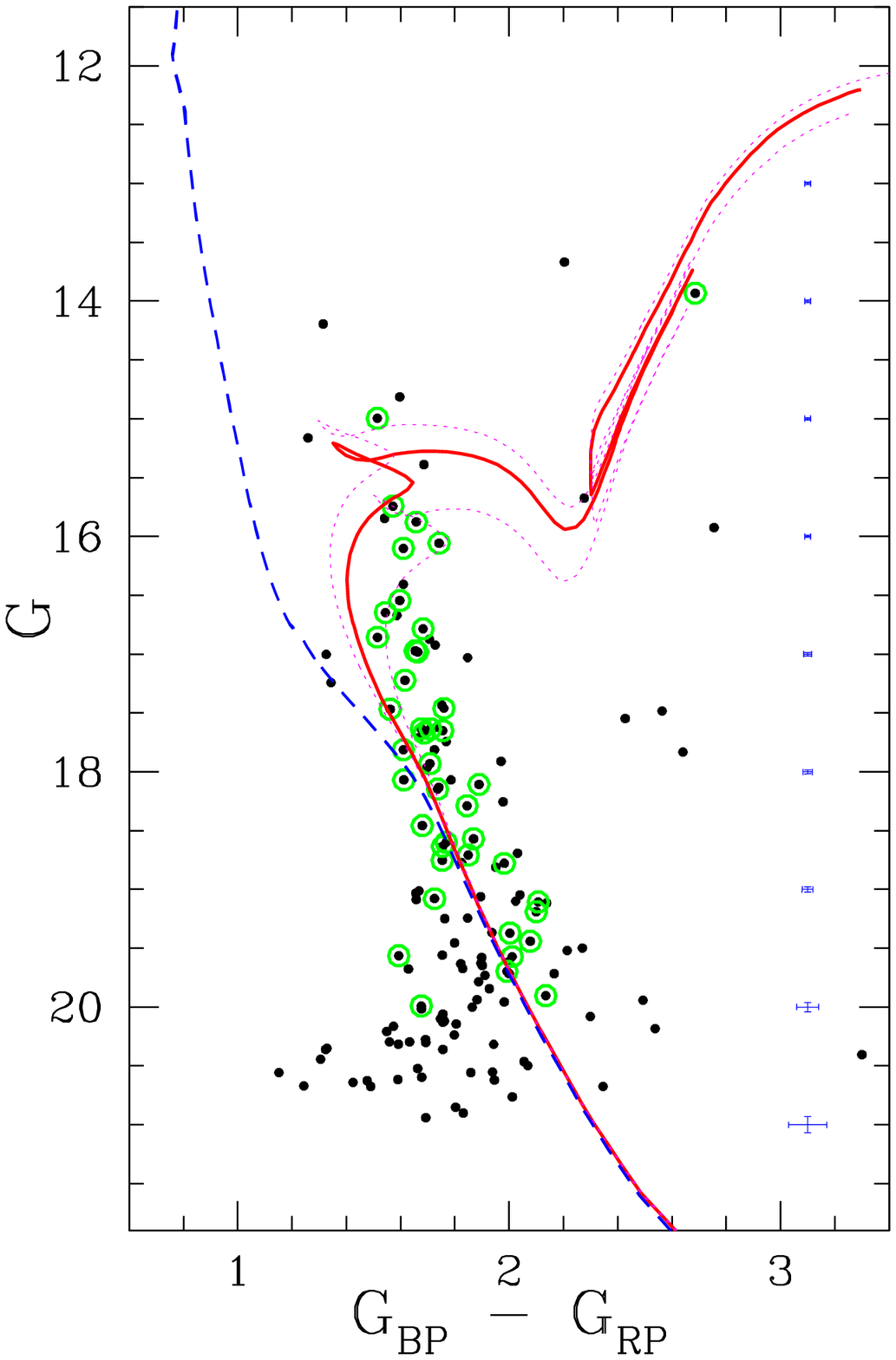}
	 \includegraphics[width=0.33\textwidth, angle=0]{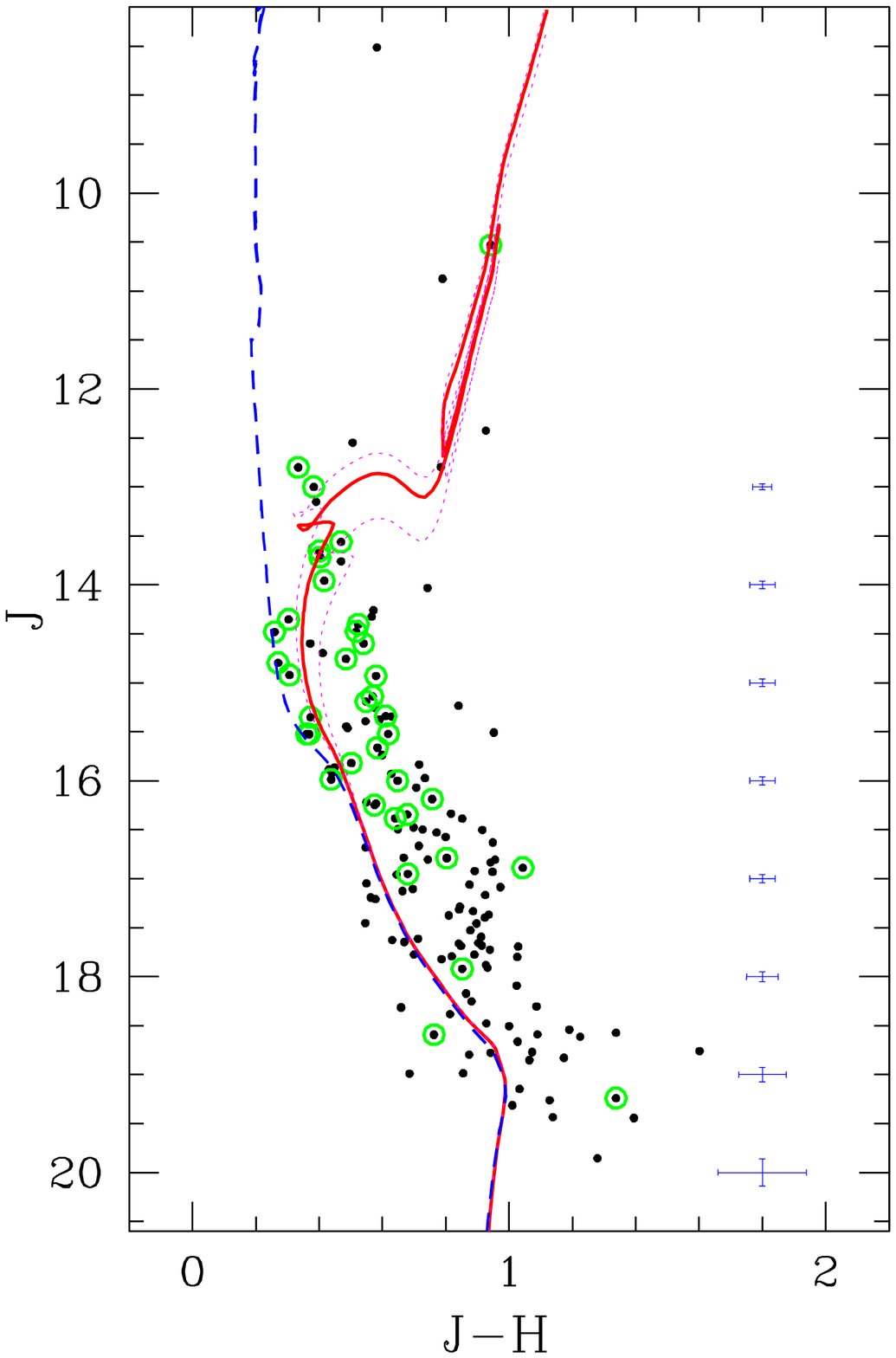}
		\caption {The CMDs generated from PS1 (left panel), $Gaia$ (middle panel), 
		and TIRCAM2 (right panel) data for the stars  in the cluster region.
		Symbols are as same as in Fig. \ref{fgy1}.
		The blue dashed and solid red curves denote ZAMS and isochrone of age = 0.9 Gyr, respectively,
		derived from \citet{2019MNRAS.485.5666P}.
		{  The magenta dotted curves denote the isochrone of age = 0.8 Gyr (upper curve)
		and 1.2 Gyr (lower curve). All the curves are 
		corrected for distance (3.5 kpc) and extinction ($A_V$=2.9 mag). 
	Photometric error bars are also shown in the CMDs.}
	 }
	  \label{fgy2}
	\end{figure*}

	\subsection{Distance and age of the cluster}

	Using the $Gaia$ data and  maximum likelihood procedure, \citet{2018yCat..36180093C} have recently calculated the
	parallax  and distance of this cluster as 0.195 mas and 4.5 kpc, respectively.
	However, they highlighted that the distances to clusters with 
	mean parallaxes smaller than $\sim$0.2 mas would be better constrained by a 
	Bayesian approach using priors based on an assumed density distribution of the
	Milky Way \citep[][]{2018AJ....156...58B} or photometric
	considerations \citep[e.g.,][]{2018AJ....156..145A}, or simply with more
	classical isochrone fitting methods \citep[e.g.,][]{1994ApJS...90...31P,2006AJ....132.1669S,2017MNRAS.467.2943S,2020MNRAS.492.2446P,2020ApJ...891...81P}.
	Therefore,  we have used both \citet{2018AJ....156...58B} approach and the isochrone fitting method
	 to constrain the  distance of this cluster.

	The mean distance of the identified member stars of this cluster comes out to be $3.5\pm0.9$ kpc \citep[see,][]{2018AJ....156...58B}.
	This distance estimation is different from all the previous measurements 
	of this cluster which places this cluster at a nearer distance (1.4 - 1.75 kpc, cf. Section 1)
	or a farther distance \citep[4.5 kpc,][]{2018yCat..36180093C}.
	Therefore, to further check the validity of this distance (3.5 kpc), we have used the 
	isochrone fitting method on the cluster CMD which is demonstrated successfully in many previous studies 
	\citep[cf.,][]{1994ApJS...90...31P,2006AJ....132.1669S,2017MNRAS.467.2943S,2014A&A...563A.117F,2015A&A...576A...6P,2019A&A...623A.108B,2020MNRAS.492.2446P,2020ApJ...891...81P}.
	To fit the isochrone to the CMD, we need the information of reddening in the
	direction of the cluster. The MWSC provides the reddening value for this cluster as $A_V \sim$2.9 mag.
	Recently, using the  $Gaia$ parallaxes and stellar photometry from PS1 and 2MASS,  
	\citet{2019ApJ...887...93G} have generated a three-dimensional map of dust reddening known as Bayestar dust map. 
	We have used their map to estimate the reddening towards the 
	direction of Cz3, which comes out to be similar to the MWSC estimate 
	and this value has been used in the present analyses.
	The relative extinction values in different passbands thereafter 
	have been estimated by using the relations given in \citet{2019ApJ...877..116W}.

	In Fig. \ref{fgy1}, we show the CMD of the stars located in the cluster region (black dots)
	using the PS1 data.  The identified member stars are shown with green circles. 
	{  We also show the zero-age main-sequence (ZAMS) 
	generated from  \citet{2019MNRAS.485.5666P} for solar metallicity (z=0.02), 
	corrected for reddening ($A_V \sim$2.9 mag)
	and distance of 1.75 kpc \citep[blue-dotted curve,][]{2017NewA...52...55B},
	3.5 kpc \citep[red-dashed curve,][]{2018AJ....156...58B} and 4.5 kpc \citep[purple-dashed curve,][]{2018yCat..36180093C}.}
	The index $(g-y)$ has been used here because of having a very large colour range.
	The ZAMS is visually fitted to the
	lower envelop of the distribution of member stars where the bend occurs in the MS.
	This choice of visual fitting was dictated by several factors, such as,
	distribution of binary stars, rotation and evolutionary effects
	\citep[see for details,][]{1974ASSL...41.....G,1994ApJS...90...31P}.
	{  Clearly, out of the three ZAMSs corrected for  different distance estimates, 
	the member stars are best represented by the ZAMS corrected for a distance of 3.5 kpc.
	Here, it is worthwhile to note that the visual fitting in this case is prone to large error
	as we do not have the error estimates in the extinction values and
	the cluster can have differential/anomalous reddening. 
	Therefore, we have assigned a distance of $3.5\pm0.9$ kpc to Cz3 cluster
	from the distance estimate of its member stars by \citet{2018AJ....156...58B}.}

	To derive the age  of this cluster, we have used the deep multi-wavelength data from 
	present observations, $Gaia$, and PS1, to generate
	CMDs in different colour spaces and are shown in  Fig. \ref{fgy2}.
	The CMDs  display a  well defined MS and a MS turn-off point.
	{  The ZAMS (generated from  \citet{2019MNRAS.485.5666P} for solar metallicity (z=0.02)) 
	corrected for the cluster's distance and reddening is also shown by a dashed blue curve.
	We can visually fit an isochrone of age $\sim$0.9 Gyr (red solid curve) 
	taken from  \citet{2019MNRAS.485.5666P} to the distribution of stars in the post-MS phase in all the CMDs. 
	For comparison, we have also plotted the isochrones of 0.8 Gyr and 1.2 Gyr in the CMDs corrected for 
	the distance and reddening which can be assumed as upper and lower limits of the age estimation \citep[see e.g.,][]{1994ApJS...90...31P}.}	

	\subsection{Mass function}

	Open clusters possess many favorable characteristics for
	MF studies, e.g., clusters contain an (almost) coeval set of
	stars at the same distance with the same metallicity; hence,
	difficulties such as complex corrections for stellar birth rates,
	life times, etc, associated with determining the 
	MF from field stars are automatically removed. 
	The MF is often expressed by a power law,
	$N (\log m) \propto m^{\Gamma}$ and  the slope of the MF is given as:

	\begin{equation}
	\Gamma = d \log N (\log m)/d \log m 
	\end{equation}

	where $N (\log m)$ is the number of stars per unit logarithmic mass interval.
	The MS luminosity function (LF) obtained with the help of the CMD 
	and corrected for the data incompleteness, has been converted into an MF 
	using the isochrone of \citet{2019MNRAS.485.5666P} of age $\sim$0.9 Gyr, 
	corrected for the distance and extinction. 

	To derive the MF slopes of the lower mass end of a cluster, it is important to know the completeness
	limits of the photometric data. The photometric data may be incomplete due to
	various reasons (e.g., crowding of stars, the detection limit, background nebulosity, etc). 
	To determine the completeness factor (CF) of the photometric data, 
	we used the ADDSTAR routine of DAOPHOT II. This method has
	been used by various authors 
	\citep[see][and references therein]{2007MNRAS.380.1141S,2008AJ....135.1934S,2017MNRAS.467.2943S,2020ApJ...891...81P}. 
	Briefly, the method consists of
	randomly adding artificial stars of known magnitudes and
	positions into the original frame. The frames are reduced with
	the same procedure used for the original frame. The ratio of the
	number of stars recovered to those added in each magnitude
	interval gives the CF as a function of magnitude. The
	luminosity distribution of artificial stars is chosen in such a
	way that more stars are inserted into the fainter magnitude bins.
	In all, about 15 percent of the total stars are added so that the
	crowding characteristics of the original frame do not change
	significantly \citep[see][]{1991A&A...250..324S}. 

	\begin{figure*}
	  \centering
	  \includegraphics[width=0.58\textwidth, angle=0]{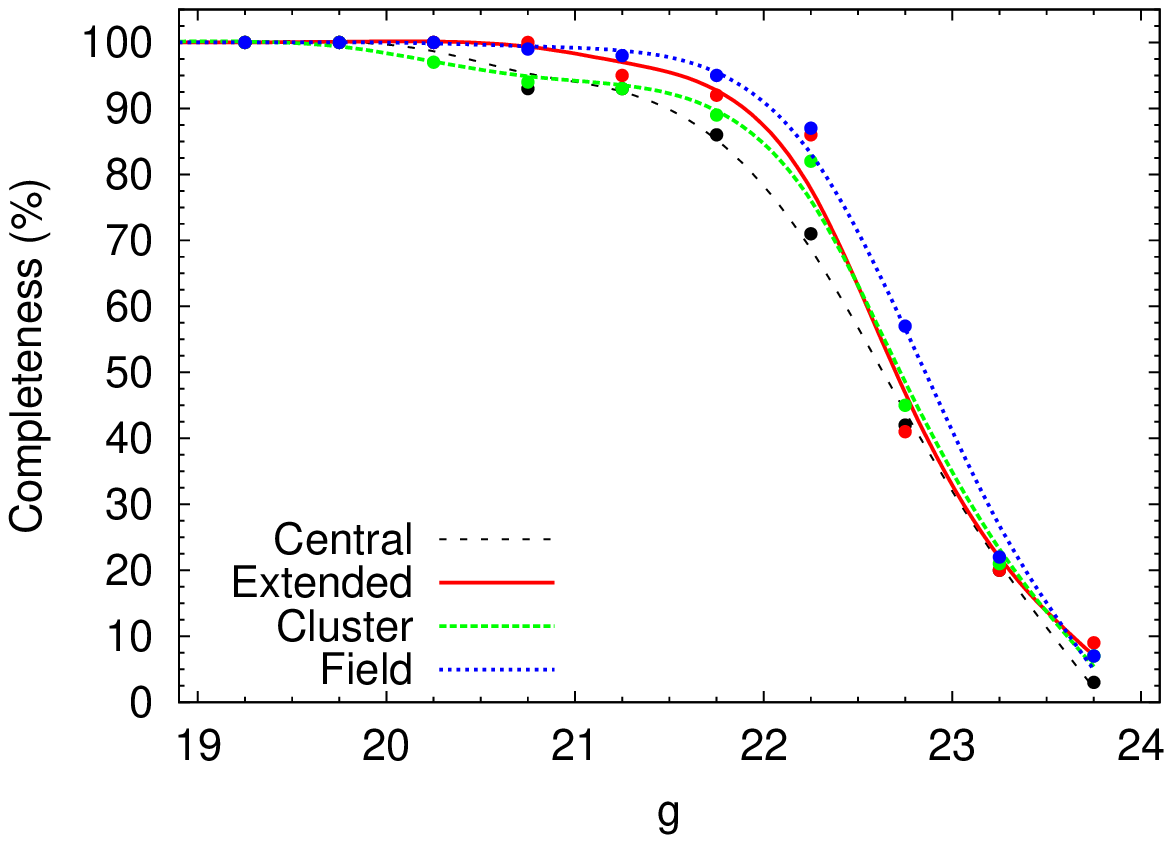}
	  \includegraphics[width=0.41\textwidth, angle=0]{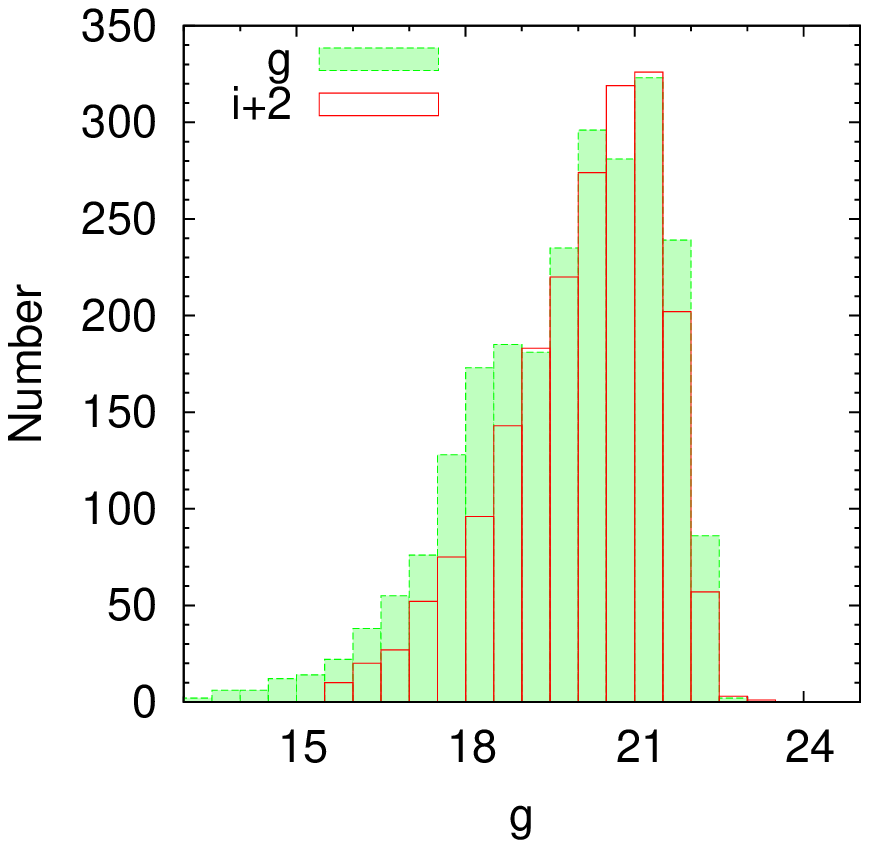}
	\caption { (Left panel): Completeness factor as a function of $g$ mag for different regions of the Cz3 cluster
		derived from the artificial star experiments ({\it ADDSTAR}, see Section 3.4 for details).
		The continuous curves are the smoothened bezier curves for the data points for completeness for 
	different regions.
		(Right Panel): Histograms for the number of stars in different magnitude bins showing the limiting magnitude and completeness factor in PS1 $g$ and $i$ bands.}
	  \label{fcftps1}
	\end{figure*}

	\begin{figure*}
	  \centering
	  \includegraphics[width=0.56\textwidth, angle=0]{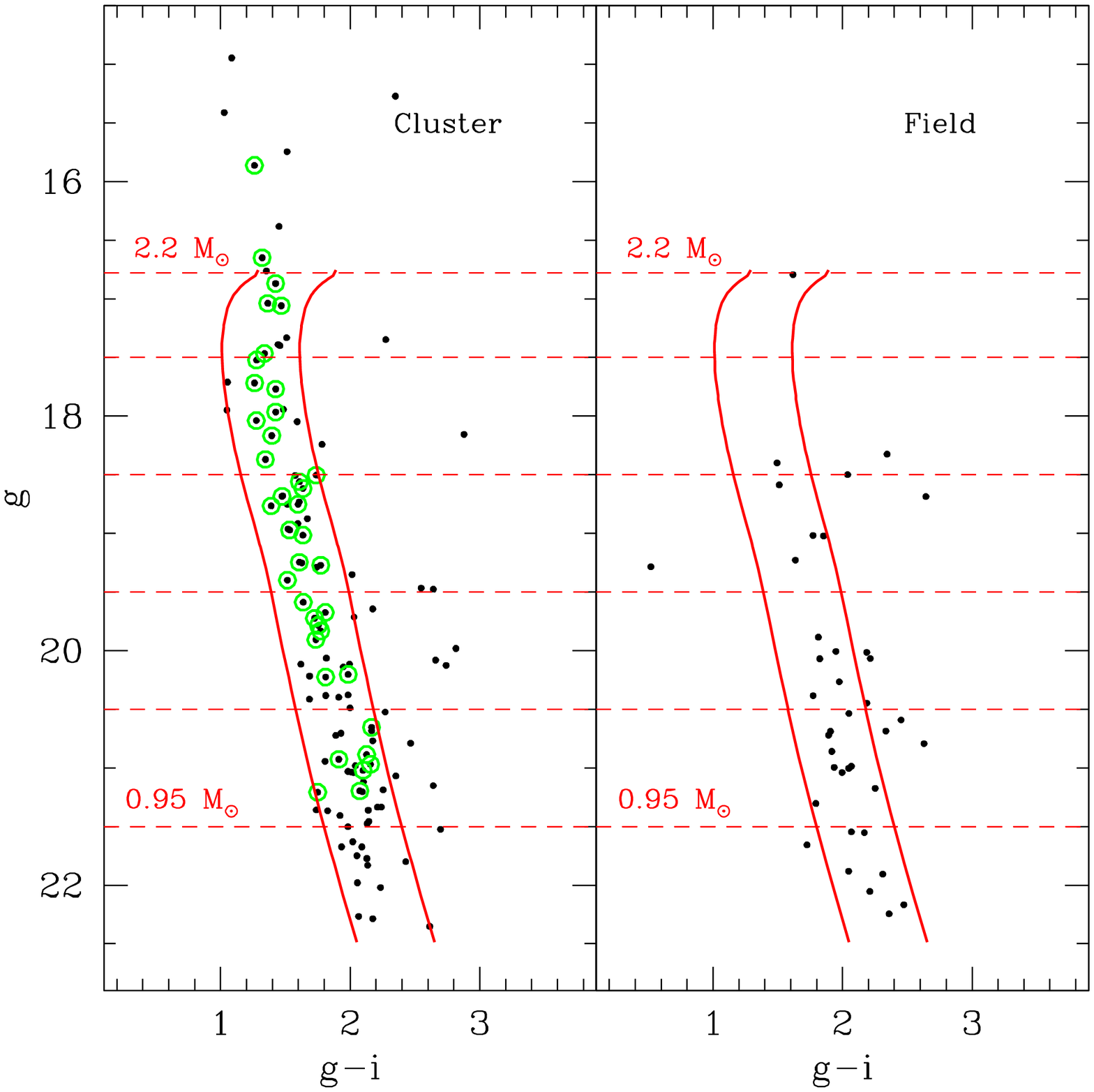}
	  \includegraphics[width=0.42\textwidth, angle=0]{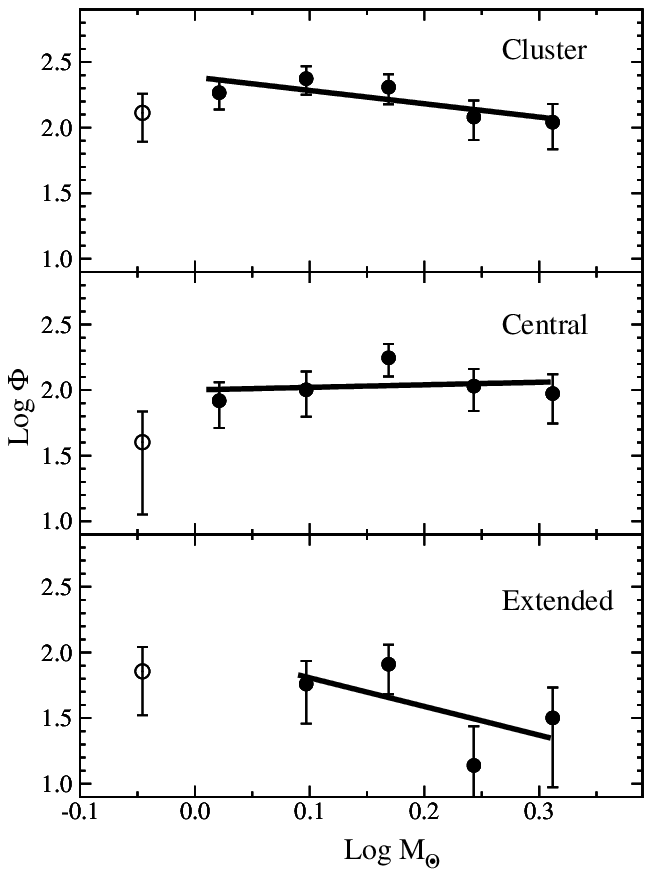}
		\caption {(Left panel): PS1 $g$ vs. $(g - i)$ CMDs for the stars  in the cluster 
		and field regions.  The symbols are same as in figure \ref{fgy1}.
		The curves denote a MS envelop created by 
		the MS isochrone of 0.9 Gyr derived from \citet{2019MNRAS.485.5666P} corrected for the
		distance (3.5 kpc) and  extinction ($A_V$=2.6 mag (left curve) and $A_V$= 3.2 mag (right curve)).
		Upper and lower horizontal dashed lines represent the MS turn-off point and 90 percent completeness limit, respectively.
		(Right panel): A plot of the MF for the cluster, central and extended regions of Cz3 using PS1 data.
	Log $\phi$ represents log($N$/dlog $m$). Open circles represent data below 50 percent completeness limit
	and are not used in the MF analysis. 
		The error bars represent $\pm\sqrt N$ errors. 
	The solid line shows a least squares fit to the MF distribution (black filled circles).
	}
	  \label{fband}
	\end{figure*}

	\begin{table}
	\centering
	\caption{\label{mfslope}The mass function slopes for two sub-regions and for the
	whole cluster region in the given mass range.}
	\begin{tabular}{@{}ccccc@{}}
	\hline
		Mass range &  \multicolumn{3}{c}{Mass Function slopes ($\Gamma$)} & \\
	 ($M_\odot$) & Central region & Extended region & Cluster region & \\
	\hline
	 $2.2-0.95^a$ & $0.21\pm0.63$  & $-2.20\pm1.90$ & $-1.01\pm0.43$ & \\
	 $2.2-0.95^b$ & $0.61\pm0.87$  & $-$ & $-$ & \\
	\hline
	\end{tabular}

	$^a$: PS1 data
	$^b$: TIRCAM2 data
	\end{table}

	\begin{figure}
	\centering
	\vbox{\includegraphics[width=0.48\textwidth, angle=0]{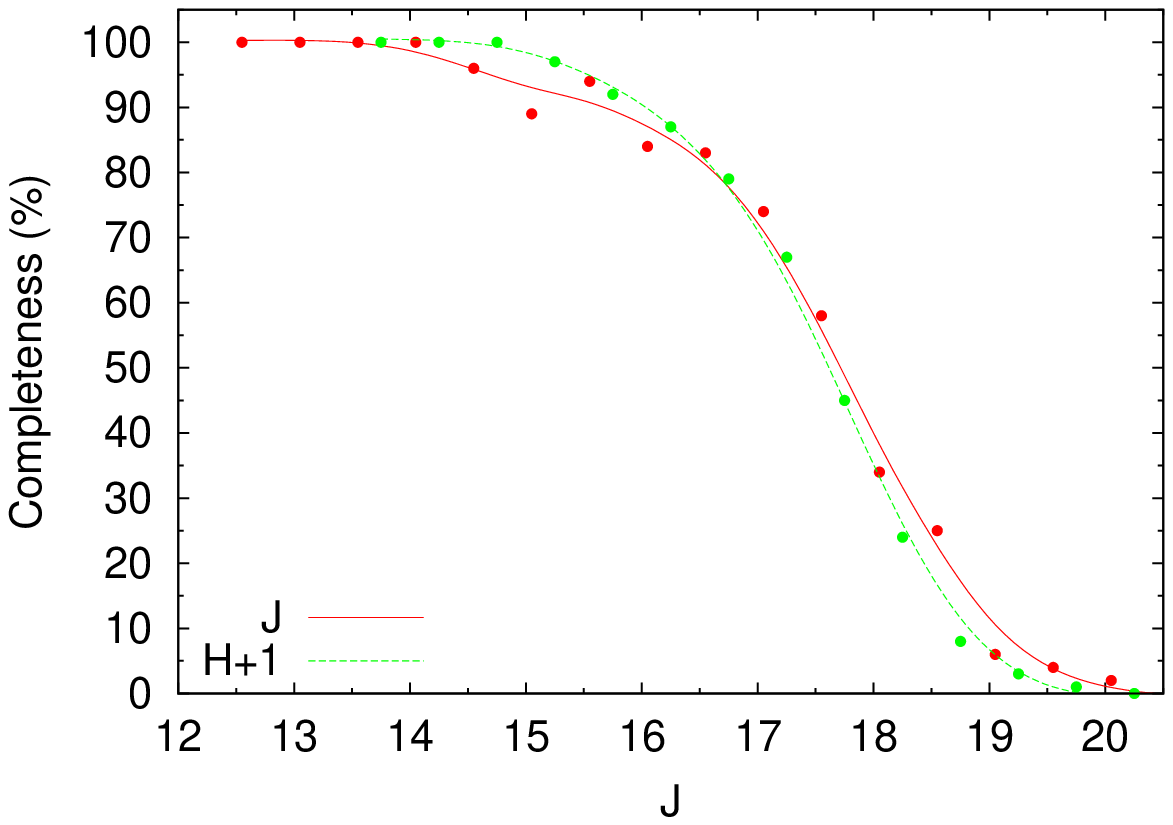}}
	\vbox{\includegraphics[width=0.48\textwidth, angle=0]{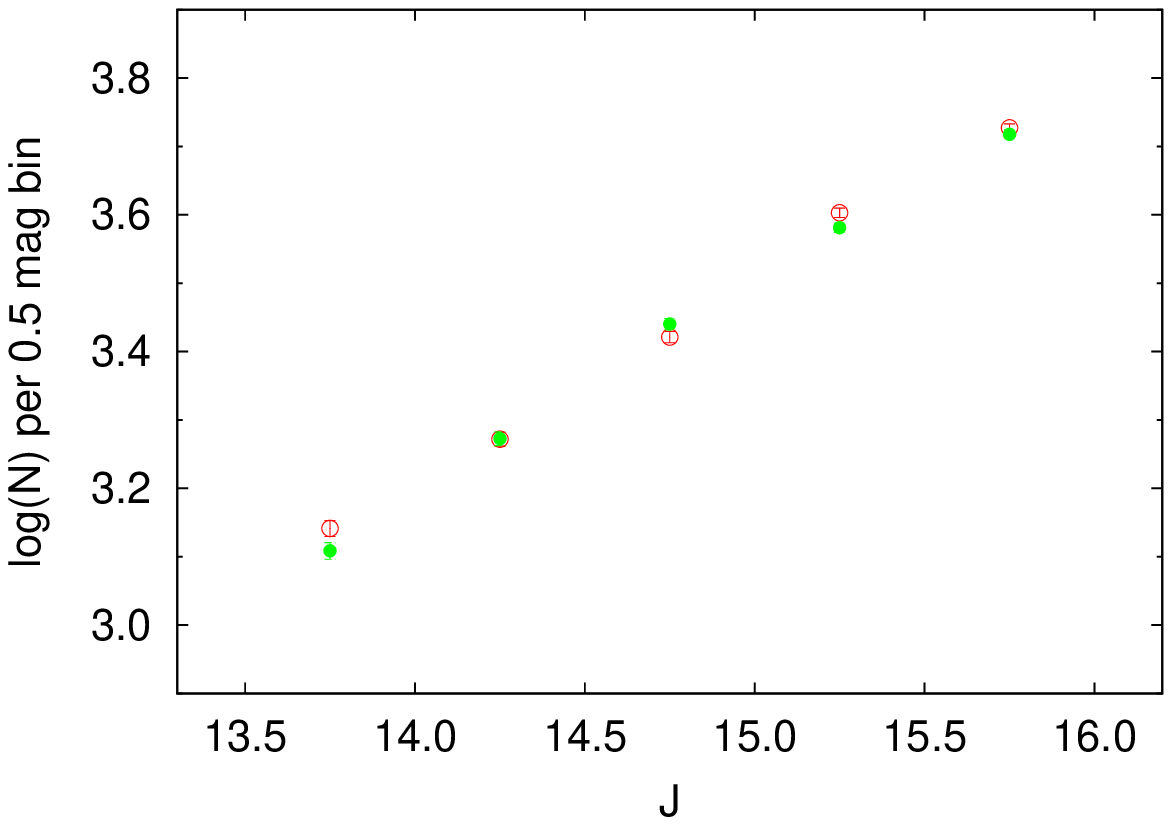}}
	\vbox{\includegraphics[width=0.48\textwidth, angle=0]{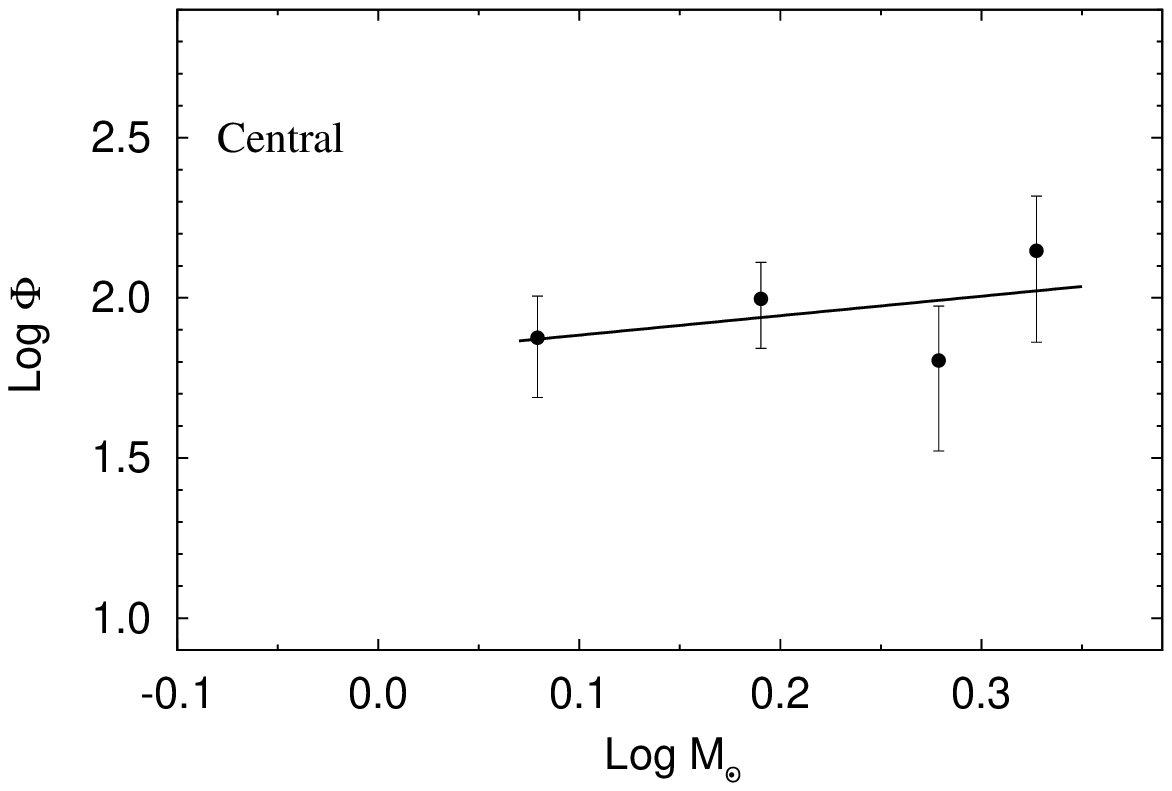}}
	\caption {(Top panel): Completeness factor as a function of $J$ magnitude
	derived from the artificial star experiments ({\it ADDSTAR}, see Section 3.4 for details)
	on the TIRCAM2 $J$ and $H$ band images.
	The $H$-band completeness factor is off-setted by the mean colour of the MS stars (i.e., 1.0 mag).
	The continuous curves are the smoothened bezier curves for the data points for completeness.
	(Middle panel): A comparison of field stars distribution generated by using a nearby reference 
	field (green filled circles)  and by the model/simulations generated by the Besan\c con model (red open circles).
	(Bottom panel): A plot of the MF for member stars in the central region of the Cz3 cluster using TIRCAM2 data.
	Log $\phi$ represents log($N$/dlog $m$). The error bars represent $\pm\sqrt N$ errors. 
	The solid line shows a least squares fit to the MF distribution (black filled circles).
	}
	\label{mf}
	\end{figure}

	We have used the deep PS1 photometric data to generate  $g$ versus $(g-i)$ CMDs
	to derive MF slopes in the cluster as well as in the central and extended regions (cf. Fig. \ref{fig:qhull}).
	Therefore, for the calculation of CF of this CMD we have used PS1 images in $g$ and $i$ bands.
	We first performed PSF photometry on these images. The detection threshold was chosen such that the
	detections are similar to that in the PS1 catalog. The instrumental magnitudes are then converted
	to the standard magnitudes by applying offsets determined by using the PS1 catalog.
	We have then added artificial stars in these images
	in such a way that they have similar geometrical locations but
	differ in $i$ brightness according to mean $(g-i)$ colours ($\sim$ 2 mag) of the MS stars.
	The minimum value of the CF of the pair thus
	obtained is used to correct the data for incompleteness.
	We have derived the CF for the central, extended, and field regions of Cz3
	and is shown as a function of $g$-band in the left panel of Fig. \ref{fcftps1}.
	As expected the incompleteness of the data increases with increasing magnitude
	and stellar crowding.
	The  photometric data is 90 percent complete upto 21.5 mag  
	in the $g$-band which corresponds to a star of mass 0.95 M$_\odot$ at the distance of the Cz3 cluster.
	The PS1 median 50 percent completeness for the $g$-band is reported to be
	23.2 mag\footnote{https://outerspace.stsci.edu/display/PANSTARRS/PS1+Photometric+Depth},
	and the present estimates are more or less similar.
	As assumed in various previous studies, the peak of the observed 
	LF can be considered as the 90 percent
	completeness limit of the data \citep[cf.][]{2003PASP..115..965E, 
	2013ApJ...778...96W, 2013MNRAS.432.3445J, 2016ApJ...822...49J, 2016AJ....151..126S,2020ApJ...891...81P,2020ApJ...889...99Z}.
	Therefore to further verify our CF values, we have constructed
	the LFs for $g$ and $i$ (off-setted by mean MS colour of 2 mag) bands
	and are shown in the right panel of Fig. \ref{fcftps1}. 
	Both the distributions are showing a peak around 21.5 mag and thus agree with our estimated 90 percent CF value.

	{  In the left panel of Fig. \ref{fband}, 
	we show the CMD for the cluster region as well as for the reference region ($\alpha_{J2000}$: $01^h04^m00^s.0$, $\delta_{J2000}$: $+62^\circ37^\prime28^{\prime \prime}$)
	having the same area.} The contamination due to field stars is greatly reduced by selecting
	a sample of stars which are located near the well-defined MS \citep[cf.][]{2008AJ....135.1934S}. 
	Therefore, we generated an envelope of $\pm 0.3$ mag around the CMD keeping in mind the distribution of member stars and
	is shown in the left panel of Fig. \ref{fband}. 
	As the MS is extended from $\sim$16.5 mag ($\sim$2.2 M$_{\odot}$) to $\sim$22.5 mag ($\sim$0.85 M$_{\odot}$), 
	the number of probable cluster members were obtained by subtracting 
	the contribution of field stars  (corrected for data incompleteness), in different magnitude bins having size of 1.0 mag
	starting from $g=16.5$ mag, 
	from the contaminated sample of MS stars (also corrected for data incompleteness).
	We have shown the MS turn-off point and 90 percent completeness limit in the left panel of Fig. \ref{fband}.
	One member stars above the MS turn-off point could be a MS blue straggler star of the cluster.
	The resultant MF distributions for the cluster as well as the central and extended
	regions are shown in the right panel of Fig. \ref{fband}.
	The  values of the MF slopes are listed in Table \ref{mfslope}.

	We have also used the present deep NIR photometry taken from  TIRCAM2 to derive MF slope of the central region of Cz3. 
	The CF is determined for $J$ versus $J-H$ CMD using the same procedure as discussed above and
	is shown in the top panel of Fig. \ref{mf}.
	To decontaminate the field star population, we have used the CMD of a nearby reference field taken from 
	the 2MASS survey (for stars having $J<16$ mag)
	and the Besan\c con Galactic model of stellar population synthesis \citep{2003A&A...409..523R,2004ApJ...616.1042O} 
	(for stars having $J>16$ mag).
	To check the accuracy of statistics of number of stars generated by the Besan\c con model, 
	we have compared the LF generated from the model with that from the 2MASS survey ($J<16$ mag) 
	in the middle panel of Fig. \ref{mf}. 
	The LFs from both methods are matching well and the resultant MF distribution is
	shown in the bottom panel of Fig. \ref{mf}. The value of MF slope for the central region of Cz3
	derived by using TIRCAM2 is listed in Table \ref{mfslope}.

	\section{Discussion}
	\label{sect:diss}

	It is well known that the internal interaction of two-body relaxation 
	due to encounters among member stars and external
	tidal forces due to the Galactic disk or giant molecular clouds can further influence 
	the structure of the clusters. 
	The vast majority of Galactic open clusters do not survive longer than a few hundred million years 
	\citep{1971A&A....13..309W,1987gady.book.....B}.
	Recently, \citet{2019arXiv190104253Y} found that the 99 percent of the cluster mass of an old open cluster Ruprecht 147 (age = 2.5 Gyr)
	was lost due to tidal interaction with the Milky way. They have found prominent leading and trailing trails of stars
	along the cluster orbit, demonstrating that the cluster is losing stars at fast pace and
	is rapidly dissolving into Galactic disc.
	Signatures of dissolution can be seen in many Galactic open clusters in the form of truncated MS in the cluster's CMD 
	\citep{1995A&AS..114..281P,1999A&A...345..485P,2005ApJ...628L.113S},
	which, in turn, produce the MF peaked closer to higher mass stars. Also, the stellar distribution in several 
	star clusters on the plane of the sky
	appears elongated or distorted  \citep{2019arXiv190104253Y}.
	We will discuss those effects in the Cz3 cluster to check its dynamical status in the following sub-sections.

	\subsection{Mass function and mass segregation}

	The higher mass stars mostly follow the Salpeter MF \citep{1955ApJ...121..161S}. At lower
	masses, the MF is less well constrained, but appears to flatten
	below 1 M$_\odot$  and exhibits fewer stars of the lowest masses 
	\citep{2002Sci...295...82K, 2003PASP..115..763C, 2015arXiv151101118L, 2016ApJ...827...52L}.
	In this study, we find a slight change of MF slope from the high to low mass end  with a turn-off
	at around 1.5 M$_\odot$ for the cluster. 
	This truncation of MF slope at a bit higher mass bins might be due to the escape of members from the cluster.
	We have also found that the MF slopes for the cluster region ($\Gamma=-1.01\pm0.43$, cf. Table \ref{mfslope}) 
	is  shallower  than the
	\citet{1955ApJ...121..161S} value i.e., $\Gamma$=-1.35, indicating the lack of low mass stars there.
	This lack of low mass  stars can be also attributed to their escape from the cluster's boundary. 
	Stars escape from the parent cluster because of a variety of processes. They can
	be divided into internal, like two body relaxation and stellar evolution
	\citep{2005A&A...429..173L, 2015ApJ...810...40D},
	and external, like tidal interaction and encounters with molecular clouds, spiral arms, 
	the Galactic disc and, in general, the interactions with the Galactic tidal field 
	\citep{1994A&AT....6...85D,2006MNRAS.371..793G,2008MNRAS.389L..28G,2019arXiv190104253Y}.

\begin{figure}
	\centering
	\includegraphics[width=0.48\textwidth, angle=0]{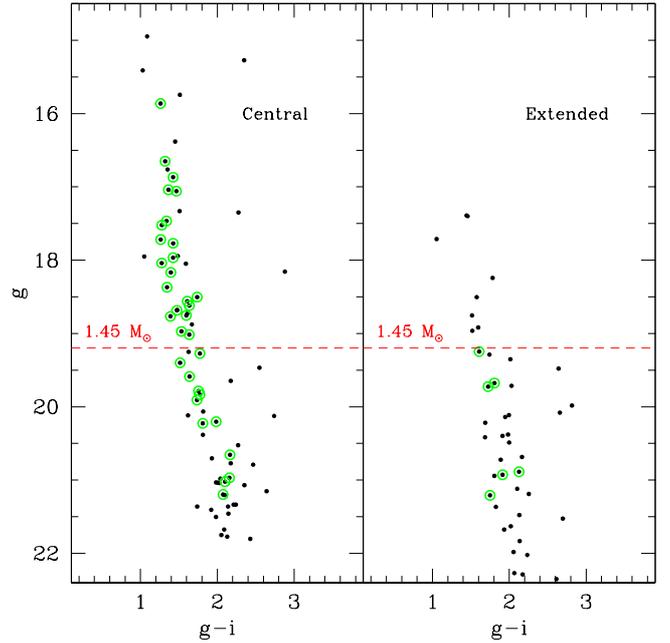}
	\caption { PS1 $g$ vs. $(g - i)$ CMDs for the stars in the  central and extended regions of the cluster (black dots).
	Member stars are shown with green circles. The horizontal dashed line corresponds to the mass of the brightest  
	star in the extended region identified as a cluster member by using PM data.}
	\label{gicn}
	\end{figure}

	In Fig. \ref{gicn}, we show the CMDs of stars in the central and extended 
	regions of the cluster. Clearly, there are many massive stars in the central region as compared
	to the extended region of the cluster.
	To investigate further, we look for the signature of mass segregation 
	by checking the change of MF slope from the central region to the outer extended region of
	this cluster, which is in fact getting steeper in the extended region (cf. Table \ref{mfslope}).
	This is an indicative of prevalent mass segregation in this cluster.
	To verify, we have also used \citet{2009MNRAS.395.1449A} method of calculating mass segregation ratio (MSR)
	as a measure to identify and quantify mass segregation in the cluster.
	This method has been updated by \citet{2011A&A...532A.119O}
	by using the geometric mean  to minimize the influence of outliers. 
	In this method, the minimal sampling tree (MST) for the $n_{MST}$ most massive stars is constructed and then
	the mean edge length $\gamma_{mp}$ is determined.
	The MST is defined as a network of branches connecting points
	such as the total length of the branches is minimized and there is no
	loops \citep{1991A&A...244...69B}. This algorithm has lately become a popular
	tool to search for clusters of stars since it is independent from
	the star's density number \citep{2009ApJS..184...18G,2014MNRAS.439.3719C,2016AJ....151..126S}.
	We then constructed the MST of the randomly selected same number of stars from the entire sample
	and determed the mean edge length $\gamma_{rand}$. Thereafter, the value of the MSR
	`$\Gamma_{MSR}$' \citep{2011A&A...532A.119O} is estimated as:

	\begin{equation}
	\Gamma_{MSR} = {{\langle \gamma^{rand}_{MST}\rangle}\over{\gamma^{mp}_{MST}}},
	\end{equation}

	and the associated standard deviation of $\Gamma_{MSR}$ is estimated as:

	\begin{equation}
	\Delta \Gamma_{MSR} = \Delta \gamma^{rand}_{MST}.
	\end{equation}

	This is done 100 times in order to obtain ${\langle \gamma^{rand}_{MST} \rangle}$.
	A value of $\Gamma_{MSR}\approx 1$ indicates that
	the most massive and the randomly selected
	are distributed in a similar manner, whereas  $\Gamma_{MSR} > 1$
	 and $\Gamma_{MSR}\ll 1$ points toward mass segreegation and inverse mass segregation, respectively \citep{2018MNRAS.473..849D}. 
	Since, the Cz3 is an old cluster with negligible differential extinction, 
	we use the magnitudes of member stars (cf. Section 3.2) as a proxy for the mass. 
	By this, we can avoid the introduction of additional uncertainties 
	of converting the observed luminosities into masses \citep{2018MNRAS.473..849D}.
	We derived  $\Gamma_{MSR} =  1.2\pm2.9$, which also suggests the effect of mass-segregation in this cluster.
	The large error in  $\Gamma_{MSR}$ is may be due to a small mass range of this old cluster.

	To decide whether mass segregation is primordial or due
	to dynamical relaxation, we have to estimate the dynamical
	relaxation time, $T_E$, the time in which the individual stars
	exchange sufficient energy so that their velocity distribution
	approaches that of a Maxwellian equilibrium
	using the method given by \citet{1987gady.book.....B}:

	\begin{equation}
	T_E = {{N}\over{8logN}}\times T_{cross}
	\end{equation}

	where $T_{cross}= {D\over\sigma_V}$ denote the crossing time, $N$ is the total number
	of stars in the region under investigation of diameter D, and $\sigma_V$ is the velocity dispersion, 
	with a typical value of 3 km s$^{-1}$ \citep{2017NewA...52...55B}. The estimated
	value of dynamical relaxation time, using the member stars (cf. Section 3.2), comes out to be $T_E \sim 10$ Myr for the Cz3 cluster.
	A comparison of cluster age (i.e. 0.9 Gyr) with its dynamical relaxation
	time indicates that the former is much greater than the
	latter, leading to the conclusion that  this cluster is dynamically relaxed and the dynamical evolution of stars could
	be the reason for the observed mass segregation in this cluster.
	One of the consequences of this mass segregation process
	is that the lowest-mass members become the most vulnerable
	to be ejected out of the system \citep[e.g., see][]{1984ApJ...284..643M}
	This stellar evaporation, with an e-folding timescale of
	$\tau_{evap} \sim 100\times T_{E}$ \citep{1982phyn.book.....S,1987gady.book.....B}, leads
	to a continuing decrease of the total mass, and hence the
	gravitational binding energy of the cluster. Any external
	disturbance, such as the tidal force from nearby giant molecular
	clouds or star clusters, passages through Galactic spiral arms or
	disks, or a shear force by Galactic differential rotation, exacerbates 
	the disintegration of the cluster. As $\tau_{evap}$ comes out to be
	$\sim$1 Gyr for Cz3 having  age$\sim$0.9 Gyr, this cluster
	can be considered as a probable candidate undergoing disintegration processes.

	\begin{figure}
	\centering
	\includegraphics[width=0.48\textwidth, angle=0]{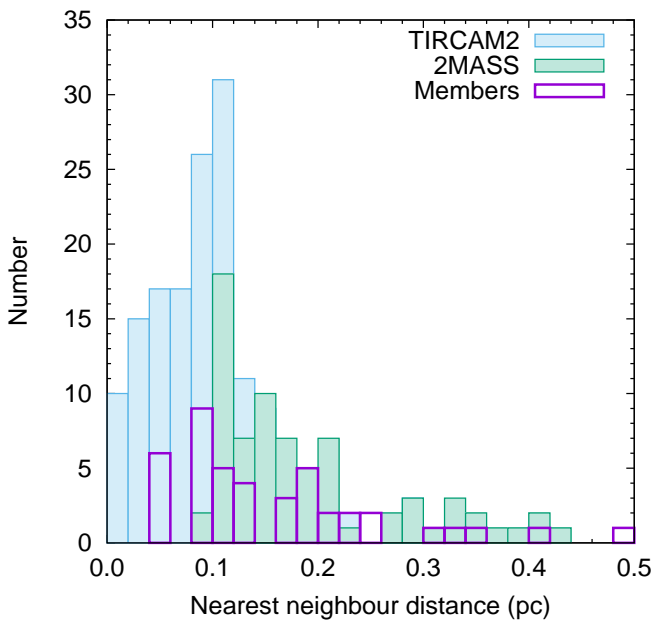}
	\caption { 
	Histogram of the nearest neighbor (NN2) lengths for the stars in the cluster region of Cz3.
	The magenta, green and blue histograms represent {\it Gaia}, 
	2MASS and TIRCAM2  data, respectively (see the text for details).
	}
	\label{nn2}
	\end{figure}

	\subsection{Morphological structure}

	By analyzing the stellar density distribution morphology, observational analyses
	can address the link between star formation, gas expulsion, and
	the dynamics of the clusters, as well as how these processes
	guide the evolution of clusters \citep{2005ApJ...632..397G}.
	From the isodensity contours  (cf. Section 3.1), we have found that this cluster is showing elongated morphology.
	{  We will now derive the Q parameter of the sample of cluster members identified in the Section 3.2}.
	The Q parameter is generally used to distinguish between clusters with a central density concentration and 
	hierarchical clusters with a fractal substructure  \citep[cf.][]{2004MNRAS.348..589C,2009MNRAS.392..341C}, to quantify the structure
	of this cluster.
	The Q parameter ($Q = { {\bar{l}_{MST}}\over{\bar{s}} }$) combines the normalized correlation length $\bar{s}$
	i.e. the mean distance between all stars, and the normalized mean
	edge length $\bar{l}_{MST}$ derived from the MST. 
	The MST mean branches length `$l_{MST}$' is calculated directly from the MST total length
	divided  by the number of branches `N' minus 1. Then, it is normalized
	by $\sqrt{A_{cluster}/N}$.  In a similar way, the mean separation
	between points `$s$' is normalized by $R_{cluster}$.
	Using the normalized values $\bar{l}_{MST}$ and $\bar{s}$, the Q parameter becomes independent 
	from the cluster size \citep{2006A&A...449..151S}. 
	According to \citet{2004MNRAS.348..589C}, both $\bar{l}_{MST}$ and
	$\bar{s}$ values decrease as the degree of radial concentration increases (or
	the level of hierarchy decreases). However, $\bar{s}$ decreases faster than
	 $\bar{l}_{MST}$. This way, a group of $(x_N, y_N )$ points distributed radially will
	have a high Q value (Q $>$ 0.8) while clusters with a more fractal
	distribution will have a low Q value (Q $<$ 0.8) \citep{2004MNRAS.348..589C,2014MNRAS.439.3719C}. 
	We have found Q=0.75, for the member stars of the Cz3 cluster (cf. Section 3.2) 
	in the magnitude range $15>G>19$ mag,
	indicating fractal distribution of stars in this cluster.
	This is in agreement with our isodensity contour structures having two peaks in the cluster region.

	The dynamical evolution of stars in a region can also be
	analyzed by deriving the typical spacing between them and
	comparing this spacing to the Jean's fragmentation scale for a
	self-gravitating medium with thermal pressure \citep{1983ASSL..106..325G}. 
	We measured the projected distance from each
	star to its nearest star neighbor (NN2)
	and plotted their histograms with a bin size of 0.02 pc
	in Fig. \ref{nn2}, for the stars in the cluster region from the TIRCAM2 as well as from the 
	2MASS data and, for the {\it Gaia} identified members stars (cf. Section 3.2).
	The TIRCAM2 sample shows a smaller NN2 distance distribution as compared to 2MASS data, may be due to a 
	better resolution of the present TIRCAM2 observations, but, all the histograms are showing
	a major peak in their distribution at $\sim$0.1 pc with a long tail at larger distances.
	\citet{2009ApJS..184...18G}, in their study of NN2 spacing's of stellar sources in 36 young star-forming clusters,
	found a well-defined peak at 0.02-0.05 pc and a tail extending
	to the spacings of 0.2 pc or greater. 
	Recently, \citet{2016AJ....151..126S} also found a well-defined single peak around 0.03 pc with extended
	spacing upto 1 pc in their histogram of NN2 spacings of stellar sources
	in a sample of eight bright-rimmed clouds (BRCs). 
	NN2 spacings of stellar sources in young star-forming regions having a well-defined single peak at smaller spacing (0.03-0.05 pc)
	indicate a significant degree of Jean's fragmentation, whereas in the Cz3 cluster case, which is comparatively an older cluster,
	the peak of similar distribution at a larger distance ($\sim$0.1 pc) indicates a role of stellar dynamics in the structural evolution
	of this cluster.

	We have also calculated the mean ($\sim$17 stars/pc$^2$) and peak ($\sim$100 stars/pc$^2$) 
	stellar densities in this cluster using the position of members stars of this cluster (cf. Section 3.2)
	by using NN based technique discussed in Section 3.1.
	\citet{2014MNRAS.439.3719C} and \citet{2009ApJS..184...18G},  have 
	derived mean  stellar densities as 33 and 60 stars/pc$^2$ and the peak  stellar densities as 150 and 200 stars/pc$^2$, respectively, 
	for their samples of embedded clusters.
	Recently, \citet{2016AJ....151..126S} 
	also found similar values for the mean ($\sim$60 stars/pc$^2$) and peak  ($\sim$150 stars/pc$^2$) stellar densities 
	in the  core region of a sample of eight BRCs.

	The tidal radius of a star cluster in the solar neighborhood 
	is computed by using the equation given by \citet{1998MNRAS.299..955P}:

	\begin{equation}
		r_t = \left\{{{GM_C}\over{2(A-B)^2}}\right\}^{1/3}
	\end{equation}

	where $G$ is the gravitational constant, $M_C$ is the total
	mass of the cluster, and $A$ and $B$ are the Oort constants, 
	$A = 15.3\pm 0.4$ kms$^{-1}$ kpc$^{-1}$, $B = −11.9\pm 0.4$ kms$^{-1}$ kpc$^{-1}$ \citep{2017MNRAS.468L..63B}. 
	Upto the completeness limit of the present photometric data i.e. 0.95 M$_\odot$, 
	the mass of the cluster is estimated as $M_C$ = 89 M$_\odot$ {  from the MF distribution of the cluster stars
	as derived from the LF of the member stars (cf. Section 3.4)}. 
	The corresponding tidal radius comes out to be $6.4 \pm0.2$ pc.
	This is a good approximation even if we have missed 50 percent of the cluster mass in
	the lower mass bins due to data incompleteness. The tidal radius will not change much 
	as the resultant tidal radius then will be $\sim$8 pc.
	Therefore, we can conclude that the tidal radius ($r_t \sim$6.4 pc) 
	of this cluster is much larger than the present estimate of 
	the cluster radius ($R_{cluster}\sim$1.2 pc).   
	Usually, the cluster boundary is taken as  one tidal radius,
	as in the case of Blanco 1 \citep{2020ApJ...889...99Z} or Coma Bernices \citep{2019ApJ...877...12T}.
	But, there are some studies  \citep[see e.g.,][]{2019A&A...621L...2R}, in which
	the cluster boundary of 2 tidal radii was adopted.
	Also the Cz3 cluster is showing elongated structure and
	the extended tail like structure might be caused by tidal forces from a nearby
	massive object, by disk crossing, or by differential rotation in the disk. 
	Tidal stripping depends on the mass and also on
	the age of a cluster, e.g., the ratio of the members inside the
	tidal radius to those outside for a young ($\sim100$ Myr) cluster Blanco 
	is reported to be 3 \citep{2020ApJ...889...99Z}.
	Whereas, for a cluster with age $\sim700$ Myr, Coma Berenices,
	this ratio comes out to be 0.6 \citep{2019ApJ...877...12T}. 
	For an old ($\sim800$ Myr) but a massive
	cluster, Hyades,  this ratio is $\sim1.1$ \citep{2019A&A...621L...2R}. 

	Therefore, the observed  small size of this old  ($\sim0.9$ Gyr) cluster looks like 
	the remain of a cluster which has already lost many of its member stars.
	The shallow MF slope, signature of mass-segregation, low density/large separation of stars, 
	elongated and distorted morphologies
	of this dynamically relaxed cluster ($T_E=10$ Myr), indicate that the Cz3 ($age=0.9$ Gyr) is 
	a loosely bound old cluster which may be under the process of dissolution under the 
	influence of external tidal interactions.

	\section{Summary and Conclusion}
	\label{sect:conclusion}

	We have performed a detailed analysis of the Cz3 open cluster using deep NIR observations taken from TIRCAM2 on 3.6m DOT
	along with the recently available high quality PM data from the {\it Gaia} DR2 and deep photometric data from PS1.
	We have investigated the structure of this cluster, determined the membership probability of stars
	in the cluster region, derived the fundamental parameters of the cluster,
	and studied the  MF and mass segregation in this cluster.
	The main results of this study can be summarized as follows:

	\begin{itemize}

	\item
	We have derived the structural parameters of this cluster by using isodensity contours 
	and found that the Cz3  cluster is showing an elongated morphology.
	The area of this cluster from PS1 data is estimated as 4.35 arcmin square which corresponds to a circular radius of 1.18 arcmin (1.2 pc).
	The core radius of this cluster is found to be 30 arcsec ($\sim$0.5 pc).

	\item
	Using {\it Gaia} DR2 data, 45 stars were marked as highly probable cluster members.
	We have estimated the distance of this cluster both using parallax of member stars and the CMD fitting technique
	and found that the cluster is located at a distance of $3.5\pm0.9$ kpc.
	We have also estimated the age of this cluster as $0.9^{+0.3}_{-0.1}$ Gyr.

	\item
	We have derived the MF slope ($\Gamma$) in the cluster region
	in the mass range $\sim$0.95$<$M/M$_\odot$$<$2.2
as $-1.01\pm0.43$, which is shallower than the value `-1.35' given by \citet{1955ApJ...121..161S}.
The cluster is showing a signature of mass-segregation and the dynamical age of this cluster is found to be
much less than the age of the cluster,
indicating that this cluster is dynamically relaxed.

\item
The Q parameter value for this cluster suggests a fractal distribution of stars in this cluster 
which is in agreement with isodensity contour structures showing elongated morphology with two peaks.
We have found a major peak in the NN2 distance distribution of the stars in this cluster at $\sim$0.1 pc, which is 
larger than that of stars in young star-forming regions ($\sim$0.03-0.05 pc).

\end{itemize}

\noindent
Finally, from the observed,  small size of this old  ($\sim$0.9 Gyr) cluster as compare to its tidal radius,
shallow MF slope, signature of mass-segregation, low density/large separation of stars, elongated and 
distorted morphology, dynamical relaxation time ($T_E$=10 Myr) compared to the 
age of Cz3 (age=0.9 Gyr), we conclude that the Cz3 is a loosely bound old disintegrating 
cluster under the influence of external tidal interactions.

\section*{Acknowledgments}

{  We thank the anonymous referee for the critical comments on the manuscript, which improved its clarity.}
We thank the staff at the 3.6m DOT, Devasthal (ARIES), for their co-operation during TIRCAM2 observations.
It is pleasure to thank the members of 3.6m DOT team and IR astronomy group at TIFR for their support
during TIRCAM2 observations.
This work has made use of data from the European Space Agency (ESA) mission
{\it Gaia} (\url{https://www.cosmos.esa.int/gaia}), processed by the {\it Gaia}
Data Processing and Analysis Consortium (DPAC,
\url{https://www.cosmos.esa.int/web/gaia/dpac/consortium}). Funding for the DPAC
has been provided by national institutions, in particular the institutions
participating in the {\it Gaia} Multilateral Agreement.
This publication also makes use of data from the Two Micron All Sky Survey, which is a joint project of the University of
Massachusetts and the Infrared Processing and Analysis Center/California Institute of Technology,
funded by the National Aeronautics and Space Administration and the National Science Foundation.
{  SS  acknowledge  the support of the Department of Science  and Technology,  Government of India, under project No. DST/INT/Thai/P-15/2019.
SKG and DKO acknowledge the support of the Department of Atomic Energy, Government of India, under project No. 
12-R\&D-0200.TFR-5.02-0200.
SBP acknowledges BRICS grant DST/IMRCD/BRICS/Pilotcall/ProFCheap/2017(G) and grant DST/INT/JSPS/P/281/2018 for the present work.}

\section*{Data availability}
The data underlying this article are available in the article and in its online supplementary material.

\bibliography{cz3}{}
\bibliographystyle{mnras}




\bsp	
\label{lastpage}
\end{document}